%%%%%%%%%%%%%%%%%%%%%%%%%%%%%%%%%%%%%%%%%%%%%%%%%%%%%%%%%%%%%%%%%%
%%%%%%%%%%%%%%%%macros start here%%%%%%%%%%%%%
\newcount\driver \newcount\mgnf \newcount\tipi
\newskip\ttglue
%%cm completo
\def\TIPITOT{
\font\dodicirm=cmr12
\font\dodicii=cmmi12
\font\dodicisy=cmsy10 scaled\magstep1
\font\dodiciex=cmex10 scaled\magstep1
\font\dodiciit=cmti12
\font\dodicitt=cmtt12
\font\dodicibf=cmbx12 scaled\magstep1
\font\dodicisl=cmsl12
\font\ninerm=cmr9
\font\ninesy=cmsy9
\font\eightrm=cmr8
\font\eighti=cmmi8
\font\eightsy=cmsy8
\font\eightbf=cmbx8
\font\eighttt=cmtt8
\font\eightsl=cmsl8
\font\eightit=cmti8
\font\seirm=cmr6
\font\seibf=cmbx6
\font\seii=cmmi6
\font\seisy=cmsy6
%%%%%%%%%%%%%%%%%%%%%%%%%%%%%%%%%%%%%%%
\font\dodicitruecmr=cmr10 scaled\magstep1
\font\dodicitruecmsy=cmsy10 scaled\magstep1
\font\tentruecmr=cmr10
\font\tentruecmsy=cmsy10
\font\eighttruecmr=cmr8
\font\eighttruecmsy=cmsy8
\font\seventruecmr=cmr7
\font\seventruecmsy=cmsy7
\font\seitruecmr=cmr6
\font\seitruecmsy=cmsy6
\font\fivetruecmr=cmr5
\font\fivetruecmsy=cmsy5
%%%% definizioni per 10pt %%%%%%%%
\textfont\truecmr=\tentruecmr
\scriptfont\truecmr=\seventruecmr
\scriptscriptfont\truecmr=\fivetruecmr
\textfont\truecmsy=\tentruecmsy
\scriptfont\truecmsy=\seventruecmsy
\scriptscriptfont\truecmr=\fivetruecmr
\scriptscriptfont\truecmsy=\fivetruecmsy
%%%%% cambio grandezza %%%%%%
\def \ottopunti{\def\rm{\fam0\eightrm}% switch to 8-point type
\textfont0=\eightrm \scriptfont0=\seirm \scriptscriptfont0=\fiverm
\textfont1=\eighti \scriptfont1=\seii   \scriptscriptfont1=\fivei
\textfont2=\eightsy \scriptfont2=\seisy   \scriptscriptfont2=\fivesy
\textfont3=\tenex \scriptfont3=\tenex   \scriptscriptfont3=\tenex
\textfont\itfam=\eightit  \def\it{\fam\itfam\eightit}%
\textfont\slfam=\eightsl  \def\sl{\fam\slfam\eightsl}%
\textfont\ttfam=\eighttt  \def\tt{\fam\ttfam\eighttt}%
\textfont\bffam=\eightbf  \scriptfont\bffam=\seibf
\scriptscriptfont\bffam=\fivebf  \def\bf{\fam\bffam\eightbf}%
\tt \ttglue=.5em plus.25em minus.15em
\setbox\strutbox=\hbox{\vrule height7pt depth2pt width0pt}%
\normalbaselineskip=9pt
\let\sc=\seirm  \let\big=\eightbig  \normalbaselines\rm
\textfont\truecmr=\eighttruecmr
\scriptfont\truecmr=\seitruecmr
\scriptscriptfont\truecmr=\fivetruecmr
\textfont\truecmsy=\eighttruecmsy
\scriptfont\truecmsy=\seitruecmsy
}\let\nota=\ottopunti}
\newfam\msbfam   %per uso in \TIPITOT
\newfam\truecmr  %per uso in \TIPITOT
\newfam\truecmsy %per uso in \TIPITOT
%%%%%%%%%%%%%%%%%%%%%%%%%%%%%%%
%%cm ridotto
\def\TIPI{
\font\eightrm=cmr8
\font\eighti=cmmi8
\font\eightsy=cmsy8
\font\eightbf=cmbx8
\font\eighttt=cmtt8
\font\eightsl=cmsl8
\font\eightit=cmti8
\font\tentruecmr=cmr10
\font\tentruecmsy=cmsy10
\font\eighttruecmr=cmr8
\font\eighttruecmsy=cmsy8
\font\seitruecmr=cmr6
\textfont\truecmr=\tentruecmr
\textfont\truecmsy=\tentruecmsy
%%%%% cambio grandezza %%%%%%
\def \ottopunti{\def\rm{\fam0\eightrm}% switch to 8-point type
\textfont0=\eightrm
\textfont1=\eighti
\textfont2=\eightsy
\textfont3=\tenex \scriptfont3=\tenex   \scriptscriptfont3=\tenex
\textfont\itfam=\eightit  \def\it{\fam\itfam\eightit}%
\textfont\slfam=\eightsl  \def\sl{\fam\slfam\eightsl}%
\textfont\ttfam=\eighttt  \def\tt{\fam\ttfam\eighttt}%
\textfont\bffam=\eightbf
\def\bf{\fam\bffam\eightbf}%
\tt \ttglue=.5em plus.25em minus.15em
\setbox\strutbox=\hbox{\vrule height7pt depth2pt width0pt}%
\normalbaselineskip=9pt
\let\sc=\seirm  \let\big=\eightbig  \normalbaselines\rm
\textfont\truecmr=\eighttruecmr
\scriptfont\truecmr=\seitruecmr
%\textfont\truecmsy=\eighttruecmsy
}\let\nota=\ottopunti}
%%am
\def\TIPIO{
\font\setterm=amr7 %\font\settei=ammi7
\font\settesy=amsy7 \font\settebf=ambx7 %\font\setteit=amit7
%%%%% cambiamenti di formato %%%
\def \settepunti{\def\rm{\fam0\setterm}% passaggio a tipi da 7-punti
\textfont0=\setterm   %\textfont1=\settei
\textfont2=\settesy   %\textfont3=\setteit
%\textfont\itfam=\setteit  \def\it{\fam\itfam\setteit}
\textfont\bffam=\settebf  \def\bf{\fam\bffam\settebf}
\normalbaselineskip=9pt\normalbaselines\rm
}\let\nota=\settepunti}

%%%%%%%%% GRAFICA
%
% Inizializza le macro postscript e il tipo di driver di stampa.
% Attualmente le istruzioni postscript vengono utilizzate solo se il driver
% e' DVILASER ( \driver=0 ), DVIPS ( \driver=1) o PSPRINT (\driver=2);
% qualunque altro valore di \driver produce un output in cui le figure
% contengono solo i caratteri inseriti con istruzioni TEX (vedi avanti).
%
%\ifnum\driver=0 \special{ps: plotfile ini.pst global} \fi
%\ifnum\driver=1 \special{header=ini.pst} \fi

\newdimen\xshift \newdimen\xwidth \newdimen\yshift
%
% inserisce una scatola contenente #3 in modo che l'angolo superiore sinistro
% occupi la posizione (#1,#2)
%
\def\ins#1#2#3{\vbox to0pt{\kern-#2 \hbox{\kern#1 #3}\vss}\nointerlineskip}
%
% Crea una scatola di dimensioni #1x#2 contenente il disegno descritto in
% #4.ps; in questo disegno si possono introdurre delle stringhe usando \ins
% e mettendo le istruzioni relative nell'argomento #3.
% Il file #4.ps contiene le istruzioni postscript, che devono essere scritte
% presupponendo che l'origine sia nell'angolo inferiore sinistro della
% scatola, mentre per il resto l'ambiente grafico e' quello standard.
% #5 deve essere della forma \eq("nome simbolico").
%
% Le istruzioni postscript possono essere inserite nel file che contiene
% l'istruzione \insertplot, racchiudendole fra le istruzioni \initfig{#4}
% e \endfig; inoltre ogni riga deve cominciare con "write13<" e deve finire
% con ">". In questo modo si crea il file #4.ps relativo alla figura.
%
\def\insertplot#1#2#3#4#5{\par%
\xwidth=#1 \xshift=\hsize \advance\xshift by-\xwidth \divide\xshift by 2%
\yshift=#2 \divide\yshift by 2%
\line{\hskip\xshift \vbox to #2{\vfil%
\ifnum\driver=0 #3
%
\special{ps: plotfile #4.ps}
\fi
\ifnum\driver=1 #3 \includegraphics{#4.ps}\fi
\ifnum\driver=2 #3 \special{
\ifnum\mgnf=0 #4.ps 1. 1. scale \fi
\ifnum\mgnf=1 #4.ps 1.2 1.2 scale\fi} \special{ini.ps}
\fi }\hfill \raise\yshift\hbox{#5}}}

%%%%%%%%%%%%%%%% GRECO

\let\a=\alpha \let\b=\beta   \let\d=\delta  \let\e=\varepsilon
 \let\g=\gamma     \let\k=\kappa  \let\l=\lambda
   \let\n=\nu       \let\p=\pi  \let\ps=\psi

  \let\z=\zeta
\let\D=\Delta   \let\G=\Gamma  \let\L=\Lambda 
\let\O=\Omega \let\P=\Pi     \let\X=\Xi

%
%%%%%%%%%%%%%%%%%%%%%  Numerazione pagine
%
\def\data{\number\day/\ifcase\month\or gennaio \or febbraio \or marzo \or
aprile \or maggio \or giugno \or luglio \or agosto \or settembre
\or ottobre \or novembre \or dicembre \fi/\number\year;\,\the\time}
\newcount\pgn \pgn=1
\def\foglio{\number\numsec:\number\pgn
\global\advance\pgn by 1}
\def\foglioa{A\number\numsec:\number\pgn
\global\advance\pgn by 1}
%

%
%%%%%%%%%%%%%%%%% EQUAZIONI CON NOMI SIMBOLICI
%%%
%%% Per assegnare un nome simbolico ad una equazione basta
%%% scrivere \Eq(...) o, in \eqalignno, \eq(...) o,
%%% nelle appendici, \Eqa(...) o \eqa(...);
%%% dentro le parentesi e al posto di ... si puo' scrivere qualsiasi commento;
%%% per avere i nomi simbolici segnati a sinistra delle formule si deve
%%% dichiarare il documento come bozza, iniziando il testo con
%%% \BOZZA. Sinonimi: \Eq,\EQ,\EQS; \eq,\eqs; \Eqa,\Eqas;\eqa,\eqas.
%%% All' inizio di ogni paragrafo si devono definire il
%%% numero del paragrafo e della prima formula dichiarando
%%% \numsec=... \numfor=...  (brevetto Eckmannn).
%%% Si possono citare formule seguenti; le corrispondenze fra nomi
%%% simbolici e numeri effettivi sono memorizzate nel file \jobname.aux, che
%%% viene letto all'inizio, se gia' presente. E' possibile citare anche
%%% formule che appaiono in altri file, purche' sia presente il
%%% corrispondente file .aux; basta includere all'inizio l'istruzione
%%%           \include{nomefile}
%%%
%%%%%%%%%%%%%%%%%%%%%%%%%%%%%%%%%%%%%%%%%%%%%%%%%%%%%%%%%%%%%%%
%
\global\newcount\numsec
\global\newcount\numfor
\global\newcount\numtheo
\global\advance\numtheo by 1

\def\senondefinito#1{\expandafter\ifx\csname#1\endcsname\relax}
\def\SIA #1,#2,#3 {\senondefinito{#1#2}%
\expandafter\xdef\csname #1#2\endcsname{#3}\else
\write16{???? ma #1,#2 e' gia' stato definito !!!!} \fi}
\def\etichetta(#1){(\veroparagrafo.\veraformula)%
\SIA e,#1,(\veroparagrafo.\veraformula) %
\global\advance\numfor by 1%
\write15{\string\FU (#1){\equ(#1)}}%
\write16{ EQ #1 ==> \equ(#1) }}
\def\letichetta(#1){\veroparagrafo.\verotheo
\SIA e,#1,{\veroparagrafo.\verotheo}
\global\advance\numtheo by 1
\write15{\string\FU (#1){\equ(#1)}}
\write16{ Sta \equ(#1) == #1 }}
\def\letichettaa(#1){A.\verotheo
\SIA e,#1,{A.\verotheo}
\global\advance\numtheo by 1
\write15{\string\FU (#1){\equ(#1)}}
\write16{ Sta \equ(#1) == #1 }}
\def\tetichetta(#1){\veroparagrafo.\veraformula %%%%copy four lines
\SIA e,#1,{(\veroparagrafo.\veraformula)}
\global\advance\numfor by 1
\write15{\string\FU (#1){\equ(#1)}}
\write16{ tag #1 ==> \equ(#1)}}
\def\FU(#1)#2{\SIA fu,#1,#2 }
\def\etichettaa(#1){(A.\veraformula)%
\SIA e,#1,(A.\veraformula) %
\global\advance\numfor by 1%
\write15{\string\FU (#1){\equ(#1)}}%
\write16{ EQ #1 ==> \equ(#1) }}
\def\BOZZA{
\def\alato(##1){%
 {\rlap{\kern-\hsize\kern-1.4truecm{$\scriptstyle##1$}}}}%
\def\aolado(##1){%
 {%\vtop to \profonditastruttura
{%\baselineskip
 %\profonditastruttura\vss
 \rlap{\kern-1.4truecm{$\scriptstyle##1$}}}}}
}
\def\alato(#1){}
\def\aolado(#1){}
\def\veroparagrafo{\number\numsec}
\def\veraformula{\number\numfor}
\def\verotheo{\number\numtheo}

\def\Eq(#1){\eqno{\etichetta(#1)\alato(#1)}}
\def\eq(#1){\etichetta(#1)\alato(#1)}
\def\leq(#1){\leqno{\aolado(#1)\etichetta(#1)}}%%%%%this line for \leqno
\def\teq(#1){\tag{\aolado(#1)\tetichetta(#1)\alato(#1)}}%%%%%this line for\tag
\def\Eqa(#1){\eqno{\etichettaa(#1)\alato(#1)}}
\def\eqa(#1){\etichettaa(#1)\alato(#1)}
\def\eqv(#1){\senondefinito{fu#1}$\clubsuit$#1
\write16{#1 non e' (ancora) definito}%
\else\csname fu#1\endcsname\fi}
\def\equ(#1){\senondefinito{e#1}\eqv(#1)\else\csname e#1\endcsname\fi}
%
%%%% next six lines by paf (no responsibilities taken)
\def\Lemma(#1){\aolado(#1)Lemma \letichetta(#1)}%
\def\Lemmaa(#1){\aolado(#1)Lemma \letichettaa(#1)}%
\def\Theorem(#1){{\aolado(#1)Theorem \letichetta(#1)}}%
\def\Proposition(#1){\aolado(#1){Proposition \letichetta(#1)}}%
\def\Corollary(#1){{\aolado(#1)Corollary \letichetta(#1)}}%
\def\Remark(#1){{\noindent\aolado(#1){\bf Remark \letichetta(#1).}}}%
\def\Definition(#1){{\noindent\aolado(#1){\bf Definition
\letichetta(#1)$\!\!$\hskip-1.6truemm}}}
\def\Example(#1){\aolado(#1) Example \letichetta(#1)$\!\!$\hskip-1.6truemm}

\def\include#1{
\openin13=#1.aux \ifeof13 \relax \else
\input #1.aux \closein13 \fi}
\openin14=\jobname.aux \ifeof14 \relax \else
\input \jobname.aux \closein14 \fi
\openout15=\jobname.aux
%

%
%%%%%%%%%%%%%%% DEFINIZIONI LOCALI
%
 
 \def\\{\noindent}

\def\tende#1{\vtop{\ialign{##\crcr\rightarrowfill\crcr
              \noalign{\kern-1pt\nointerlineskip}
              \hskip3.pt${\scriptstyle #1}$\hskip3.pt\crcr}}}
\def\otto{{\kern-1.truept\leftarrow\kern-5.truept\to\kern-1.truept}}
 \def\Z{{\bf Z}}\def\R{{\bf R}}
\def\mbox{\hbox}

\def\={{\equiv}}\def\ch{{\chi}}
\def\initfiat#1#2#3{
\mgnf=#1
\driver=#2
\tipi=#3
\ifnum\tipi=0\TIPIO \else\ifnum\tipi=1 \TIPI\else \TIPITOT\fi\fi
%\ifnum\driver=0 \special{ps: plotfile ini.pst global} \fi
%\ifnum\driver=1 \special{header=ini.pst} \fi
%
%%%%%%%%%%%%%%% FORMATO
%
\ifnum\mgnf=0
\magnification=\magstep0 \hoffset=0.cm
\voffset=-1truecm\hoffset=-.5truecm\hsize=16.5truecm \vsize=25.truecm
\baselineskip=14pt  % plus0.1pt minus0.1pt
\parindent=12pt
\lineskip=4pt\lineskiplimit=0.1pt      \parskip=0.1pt plus1pt
\def\ds{\displaystyle}\def\st{\scriptstyle}\def\sst{\scriptscriptstyle}
\font\seven=cmr7
\fi
\ifnum\mgnf=1
\magnification=\magstep1
\hoffset=0.cm
\voffset=-1truecm
\hoffset=-.5truecm
\hsize=16.5truecm
\vsize=25truecm
\baselineskip=12pt
% plus0.1pt minus0.1pt
\parindent=12pt
\lineskip=4pt\lineskiplimit=0.1pt\parskip=0.1pt plus1pt
\def\ds{\displaystyle}\def\st{\scriptstyle}\def\sst{\scriptscriptstyle}
\font\seven=cmr7
\fi
\setbox200\hbox{$\scriptscriptstyle \data $}
}
%%%%%%%%%%%end of Gallavotti's macros%%%%%%%%%
%%%%%%%%%%%inizialization%%%%%%%%%%%
%%%%%%put % in front of \BOZZA to remove labels on the left%%%%%%%%%%%
\initfiat {1}{1}{2}
%\BOZZA
%\input amssym.def%
%%%%%%%%amssym.def included here%%%%%%%%%%%%
%
%%%%%%%%%%%%%%%%%%%%%%%%%%%%%%%%%%%%%%%%%%%%%%%%%%%%%%%%%%%%%%%%%%%%%%%%
\expandafter\ifx\csname amssym.def\endcsname\relax \else\endinput\fi
%
%  Store the catcode of the @ in the csname so that it can be restored later.
\expandafter\edef\csname amssym.def\endcsname{%
       \catcode`\noexpand\@=\the\catcode`\@\space}
%  Set the catcode to 11 for use in private control sequence names.
\catcode`\@=11
%
%  Include all definitions related to the fonts msam, msbm and eufm, so that
%  when this file is used by itself, the results with respect to those fonts
%  are equivalent to what they would have been using AMS-TeX.
%  Most symbols in fonts msam and msbm are defined using \newsymbol;
%  however, a few symbols that replace composites defined in plain must be
%  defined with \mathchardef.
%
\def\undefine#1{\let#1\undefined}
\def\newsymbol#1#2#3#4#5{\let\next@\relax
 \ifnum#2=\@ne\let\next@\msafam@\else
 \ifnum#2=\tw@\let\next@\msbfam@\fi\fi
 \mathchardef#1="#3\next@#4#5}
\def\mathhexbox@#1#2#3{\relax
 \ifmmode\mathpalette{}{\m@th\mathchar"#1#2#3}%
 \else\leavevmode\hbox{$\m@th\mathchar"#1#2#3$}\fi}
\def\hexnumber@#1{\ifcase#1 0\or 1\or 2\or 3\or 4\or 5\or 6\or 7\or 8\or
 9\or A\or B\or C\or D\or E\or F\fi}
\font\tenmsa=msam10
\font\sevenmsa=msam7
\font\fivemsa=msam5
\newfam\msafam
\textfont\msafam=\tenmsa
\scriptfont\msafam=\sevenmsa
\scriptscriptfont\msafam=\fivemsa
\edef\msafam@{\hexnumber@\msafam}
\mathchardef\dabar@"0\msafam@39
\def\dashrightarrow{\mathrel{\dabar@\dabar@\mathchar"0\msafam@4B}}
\def\dashleftarrow{\mathrel{\mathchar"0\msafam@4C\dabar@\dabar@}}

\def\ulcorner{\delimiter"4\msafam@70\msafam@70 }
\def\urcorner{\delimiter"5\msafam@71\msafam@71 }
\def\llcorner{\delimiter"4\msafam@78\msafam@78 }
\def\lrcorner{\delimiter"5\msafam@79\msafam@79 }
\def\yen{{\mathhexbox@\msafam@55 }}
\def\checkmark{{\mathhexbox@\msafam@58 }}
\def\circledR{{\mathhexbox@\msafam@72 }}
\def\maltese{{\mathhexbox@\msafam@7A }}
\font\tenmsb=msbm10
\font\sevenmsb=msbm7
\font\fivemsb=msbm5
\newfam\msbfam
\textfont\msbfam=\tenmsb
\scriptfont\msbfam=\sevenmsb
\scriptscriptfont\msbfam=\fivemsb
\edef\msbfam@{\hexnumber@\msbfam}
\def\Bbb#1{{\fam\msbfam\relax#1}}
\def\widehat#1{\setbox\z@\hbox{$\m@th#1$}%
 \ifdim\wd\z@>\tw@ em\mathaccent"0\msbfam@5B{#1}%
 \else\mathaccent"0362{#1}\fi}
\def\widetilde#1{\setbox\z@\hbox{$\m@th#1$}%
 \ifdim\wd\z@>\tw@ em\mathaccent"0\msbfam@5D{#1}%
 \else\mathaccent"0365{#1}\fi}
\font\teneufm=eufm10
\font\seveneufm=eufm7
\font\fiveeufm=eufm5
\newfam\eufmfam
\textfont\eufmfam=\teneufm
\scriptfont\eufmfam=\seveneufm
\scriptscriptfont\eufmfam=\fiveeufm

%
%  Restore the catcode value for @ that was previously saved.
\csname amssym.def\endcsname
%
%
%%%%%%%%%%%%end of amssym.def%%%%%%%%
%
%%%%%%%%%%%%%%%%%%extra definitions already in yau's file%%%%%%%%
%
\def\sqr#1#2{{\vcenter{\vbox{\hrule height.#2pt
     \hbox{\vrule width.#2pt height#1pt \kern#1pt
   \vrule width.#2pt}\hrule height.#2pt}}}}
\def\qed{ $\mathchoice\sqr64\sqr64\sqr{2.1}3\sqr{1.5}3$}

\def\11{\hbox{l}\!\!\!1\,}

\font\tenib=cmmib10
\newfam\mitbfam
\textfont\mitbfam=\tenib
\scriptfont\mitbfam=\seveni
\scriptscriptfont\mitbfam=\fivei

 %bold eta
%above ceta because bold eta should not be beta

%\def\bomicron{{\mitb\mathchar"7122}}
%above, not omicron, sort of script e lower case

%\baselineskip7mm
\def\and{ \hbox{ and } }

\def\pt{\partial}
\def\l{\lambda}
\def\L{\Lambda}
\def\e{\varepsilon}
\def\a{\alpha}

\def\d{\delta}
\def\g{\gamma}

\def\O{\Omega}
\def\n{\nabla}

\def\t{\theta}

\def\to{\rightarrow}

\def\frac{\over}
\def\\{\cr}
\def\ref{}

\hbox{}
\vfill
\baselineskip12pt
\overfullrule=0in

\def\a{\alpha}

\def\d{\delta}

\def\z{\zeta}
\def\I{\Bbb I}

\def\text{\hbox}
\def\\\ {\cr}

\def\ve{\varepsilon}
%
%%%%%%%%References macros start here%%%%%%%%%%
%
\catcode`\X=12\catcode`\@=11
\def\n@wcount{\alloc@0\count\countdef\insc@unt}
\def\n@wwrite{\alloc@7\write\chardef\sixt@@n}
\def\n@wread{\alloc@6\read\chardef\sixt@@n}
\def\crossrefs#1{\ifx\alltgs#1\let\tr@ce=\alltgs\else\def\tr@ce{#1,}\fi
   \n@wwrite\cit@tionsout\openout\cit@tionsout=\jobname.cit
   \write\cit@tionsout{\tr@ce}\expandafter\setfl@gs\tr@ce,}
\def\setfl@gs#1,{\def\@{#1}\ifx\@\empty\let\next=\relax
   \else\let\next=\setfl@gs\expandafter\xdef
   \csname#1tr@cetrue\endcsname{}\fi\next}
\newcount\sectno\sectno=0\newcount\subsectno\subsectno=0\def\r@s@t{\relax}
\def\resetall{\global\advance\sectno by 1\subsectno=0
  \gdef\firstpart{\number\sectno}\r@s@t}
\def\resetsub{\global\advance\subsectno by 1
   \gdef\firstpart{\number\sectno.\number\subsectno}\r@s@t}
\def\v@idline{\par}\def\firstpart{\number\sectno}
\def\l@c@l#1X{\firstpart.#1}\def\gl@b@l#1X{#1}\def\t@d@l#1X{{}}
\def\m@ketag#1#2{\expandafter\n@wcount\csname#2tagno\endcsname
     \csname#2tagno\endcsname=0\let\tail=\alltgs\xdef\alltgs{\tail#2,}%
  \ifx#1\l@c@l\let\tail=\r@s@t\xdef\r@s@t{\csname#2tagno\endcsname=0\tail}\fi
   \expandafter\gdef\csname#2cite\endcsname##1{\expandafter
 %the following line was replaced by the subseqent one, DNA 7/6/89
  %  \ifx\csname#2tag##1\endcsname\relax?\else\csname#2tag##1\endcsname\fi
     \ifx\csname#2tag##1\endcsname\relax?\else{\rm\csname#2tag##1\endcsname}\fi
    \expandafter\ifx\csname#2tr@cetrue\endcsname\relax\else
     \write\cit@tionsout{#2tag ##1 cited on page \folio.}\fi}%
   \expandafter\gdef\csname#2page\endcsname##1{\expandafter
     \ifx\csname#2page##1\endcsname\relax?\else\csname#2page##1\endcsname\fi
     \expandafter\ifx\csname#2tr@cetrue\endcsname\relax\else
     \write\cit@tionsout{#2tag ##1 cited on page \folio.}\fi}%
   \expandafter\gdef\csname#2tag\endcsname##1{\global\advance
     \csname#2tagno\endcsname by 1%
   \expandafter\ifx\csname#2check##1\endcsname\relax\else%
\fi%      \immediate\write16{Warning: #2tag ##1 used more than once.}\fi
   \expandafter\xdef\csname#2check##1\endcsname{}%
   \expandafter\xdef\csname#2tag##1\endcsname
     {#1\number\csname#2tagno\endcsnameX}%
   \write\t@gsout{#2tag ##1 assigned number \csname#2tag##1\endcsname\space
      on page \number\count0.}%
   \csname#2tag##1\endcsname}}%
\def\m@kecs #1tag #2 assigned number #3 on page #4.%
   {\expandafter\gdef\csname#1tag#2\endcsname{#3}
   \expandafter\gdef\csname#1page#2\endcsname{#4}}
\def\re@der{\ifeof\t@gsin\let\next=\relax\else
    \read\t@gsin to\t@gline\ifx\t@gline\v@idline\else
    \expandafter\m@kecs \t@gline\fi\let \next=\re@der\fi\next}
\def\t@gs#1{\def\alltgs{}\m@ketag#1e\m@ketag#1s\m@ketag\t@d@l p
    \m@ketag\gl@b@l r \n@wread\t@gsin\openin\t@gsin=\jobname.tgs \re@der
    \closein\t@gsin\n@wwrite\t@gsout\openout\t@gsout=\jobname.tgs }
\outer\def\localtags{\t@gs\l@c@l}
\outer\def\globaltags{\t@gs\gl@b@l}
\outer\def\newlocaltag#1{\m@ketag\l@c@l{#1}}
\outer\def\newglobaltag#1{\m@ketag\gl@b@l{#1}}

\def\t@gsoff#1,{\def\@{#1}\ifx\@\empty\let\next=\relax\else\let\next=\t@gsoff
   \expandafter\gdef\csname#1cite\endcsname{\relax}
   \expandafter\gdef\csname#1page\endcsname##1{?}
   \expandafter\gdef\csname#1tag\endcsname{\relax}\fi\next}
\def\verbatimtags{\let\ift@gs=\iffalse\ifx\alltgs\relax\else
   \expandafter\t@gsoff\alltgs,\fi}
\catcode`\X=11 
%\catcode`\@=\active
\localtags
%%%%%%%%%%%%references macro end here%%%%%%%%%%%%%%%
%
%%%%%%%%%%%%%%%%%end of macros%%%%%%%%%%%%%%%%%%%%%%

\let\epsilon=\varepsilon
\def\ds{\displaystyle}
\def\st{\scriptstyle}
\def\sst{\scriptscriptstyle}
%%%%%%%%%%%%%%%text starts here%%%%%%%%%%%%%%%%%%%%%
%%%%%%%%%%%%%%%%%%%%%%%%%% Local macros %%%%%%%%%%%%%%%%%%%%%%%%%%
 \font\title=cmbx12\def\1{\,\rlap{\ninerm 1}\kern.15em 1}
\def\cd{{\cal D}}
\def\ce{{\cal E}} 
\def\cg{{\cal G}}
\def\ch{{\cal H}}
\def\cl{{\cal L}} \def\cle{\cl^{\rm ex}} \def\clc{\cl^{\rm c}}
\def\cq{{\cal Q}} \def\cv{{\cal V}} \def\div{\,{\rm div}}

\let\n=\eta \let\na=\nabla \let\ov=\overline \let\p=\partial
\def\ps#1{\bigl\langle #1\bigr\rangle} 

 \let\ve=\varepsilon \let\vp=\varpi
   
\def\R{\Bbb R} \def\T{\Bbb T} \def\Z{\Bbb Z} \def\E{\Bbb E} \def\P{\Bbb P}
\def\bl{\bigl} \def\br{\bigr} \def\Bl{\Bigl}  \def\Br{\Bigr}
\def\lt{\left} \def\rt{\right} \def\D{\Bbb D}
%%%%%%%%%%%%%%%%%%%%%%%%%% Local Format %%%%%%%%%%%%%%%%%%%%%%%%%%
%\magnification=\magstep1
\hoffset=0.cm \voffset=-.2truecm \hoffset=-.2truecm \hsize=16.5truecm
\vsize=24truecm \baselineskip=12pt \parindent=12pt
\lineskip=4pt\lineskiplimit=0.1pt\parskip=0.1pt plus1pt
%% comment next line to remove the printing of labels
%\BOZZA

\centerline {\title Equilibrium Fluctuations for Lattice Gases.}
\vskip 1truecm 
\centerline{\baselineskip=10pt 
  O. Benois\footnote{$^1$}{\eightrm Laboratoire Salem, 
UMR 6085, site Colbert, Universit\'e de Rouen,
    76821 Mont Saint Aignan, France}, 
  \hskip.2cm 
  R. Esposito\footnote{$^2$}{\eightrm Dipartimento di
Matematica Pura ed
    Applicata, Universit\`a di L'Aquila, 67100 Coppito, AQ, Italy}
\footnote{$ $} {\eightrm and Centro Linceo Interdisciplinare ``Beniamino
Segre'', Via della Lungara 10, 00165 Roma, Italy},
  \hskip.2cm 
  R. Marra\footnote{$^3$}{\eightrm Dipartimento di Fisica e
    Unit\`a INFM, Universit\`a di Roma Tor Vergata, Via della Ricerca
    Scientifica,}\footnote{$ $} {\eightrm00133 Roma, Italy}}
\vskip.5cm
\centerline{Universit\'e de Rouen, Universit\`a di L'Aquila, Universit\`a di Roma Tor
  Vergata}

\bigskip 
{\baselineskip = 8pt\rightskip1.4cm\leftskip 1.4cm\ottopunti
  {\noindent {\bf Abstract:}} \quad 
  The authors in a previous paper proved the hydrodynamic
  incompressible limit in $d\ge 3$ for a thermal lattice gas, namely a
  law of large numbers for the density, velocity field and energy.  In
  this paper the equilibrium fluctuations for this model are studied and
  a central limit theorem is proved for a suitable modification of the vector
fluctuation field
  $\z(t)$, whose components are the density, velocity and energy
  fluctuations fields.  We consider 
  a modified fluctuation field
  $\xi^\e(t)=\exp \{-\ve^{-1}t E\}\z^\ve$, where $E$ is the linearized
  Euler operator around the equilibrium and prove that $\xi^\e(t)$
  converges to a vector generalized Ornstein-Uhlenbeck process
  $\xi(t)$, which is formally solution of the stochastic differential
  equation $d \xi(t)= N\xi(t)dt+ B dW_t$, with $ BB^*=-2 NC$, where
  $C$ is the compressibility matrix, $N$ is a matrix whose entries are
  second order differential operators and $B$ is a mean zero Gaussian
  field. The relation $-2NC=BB^*$ is the fluctuation-dissipation
  relation.
}

\bigskip
\noindent{\ottopunti {\bf Key Words:}
Fluctuations, Stochastic Cellular Automata, 
Navier-Stokes Equation.}
\bigskip
\noindent{\ottopunti {\bf MSC2000 Classification:}
60K35, 82C22}

\vskip1.5cm 
\noindent{\bf 1. Introduction.}
\vskip .5truecm  \numsec= 1 \numfor= 1 \numtheo=1

In this paper we study the equilibrium fluctuations for the stochastic
lattice gas introduced in [BEM]. It is a model of particles with
discrete velocities jumping on the lattice: a particle with a given
velocity moves on the 3-d lattice as the asymmetric simple exclusion
process with the jump intensity chosen so to have a drift equal to its
velocity. In each site particles collide exchanging velocities in such
a way to conserve the number of particles, the momentum in each
direction and the energy. This model generalizes the one in [EMY2] to
include the case of particles with different kinetic energy.  In [BEM]
it has been proved the law of large number for this model in the
following form. We choose as initial state a Bernoulli measure with
density, momentum and energy small perturbation (of order $\e$) of
constant profiles. Then the empirical fields
$\nu_\b^\e(x,t)=\nu_\b(\ve^{-1}x, \ve^{-2}t), \b=0,\dots,4$ of
density, momentum and energy converge weakly in probability as $\e$
goes to $0$ to the solution of the hydrodynamic equations for this
model, which are the incompressible Navier-Stokes equation for the
velocity field and a diffusive equation (including the transport along
the velocity field) for the energy. The dissipative terms in these
equations are given in terms of a diffusion tensor $D^{\b,\nu}_{\a,\g},
\b,\nu=0,\dots,4,\ \a,\g=1,\dots,3$, which is expressed by the
Green-Kubo formulas.  The next natural step is to prove the space-time
central limit theorem, namely that the fluctuations fields starting
from the equilibrium state converge to a stationary multi-dimensional
Gaussian process with a given space-time covariance. Since the
macroscopic behavior of this model is very close to the real
hydrodynamics we face the main difficulty of the hydrodynamic
fluctuations: the Euler terms and the Navier-Stokes terms live on
different time scales. The same feature is responsible for the
impossibility of obtaining the compressible dissipative hydrodynamic
equations as scaling limit. In fact, the previous result on the law of
large number is true for an initial condition which is a small
perturbation of the global equilibrium. This perturbation remains
small at later times of order $\ve^{-2}$ and evolves macroscopically
according to the incompressible hydrodynamics. The case of the
fluctuations is different because a small perturbation of the
equilibrium may become very large and be of order $\ve^{-1}$ on times
of order $\ve^{-2}$. We go now in some details to explain better this
point. The fluctuation fields
%change
under diffusive scaling 
are defined by
$$
\z^\e _\b(t,G)=\ve^{3\over 2}\sum_x G_\b(\e x)
\big[I_\b(\eta_{\ve^{-2}t}(x) -\E[I_\b]\big], %\Eq(0.1)
$$
$\b=0,\cdots,4$, where $G_\b$ are suitable test functions, $\eta_t(x)$ is the
configuration in $x$ at time $t$ and
$\E$ is the equilibrium expectation. $I_\b$ are the quantities conserved by
the dynamics, total number of particles, total momentum and total
energy in $x$.  

\bigskip

At time zero the limiting fluctuation fields
$$
\lim_{\e\to 0}\z^\e _\b(0,G)=\z _\b(0,G)
$$
are jointly Gaussian with covariance
$$
\E[\z_\b(0,G)\z_\nu(0,H)]= C_{\b,\nu}\int\!d^3 x\,\ G(x)H(x).
$$
The matrix
$C=(C_{\b,\nu })$ is called the compressibility matrix. The limit is
in the in sense of weak convergence of path measures.

\medskip

It is not hard to show (it is indeed a by-product of the results and
estimates in this paper) that the equilibrium fluctuations 
%change
%on a time scale $\tau=\ve^{-1}t$ 
under Euler time scale
are trivial in the sense that they
satisfy in the limit a deterministic equation.  
This is a general feature first
showed in [GP], [FF]. More precisely, the limiting field
$\z^E=(\z_\b^E), \b=0,\cdots,4$
$$
\z^E(\tau)=\lim_{\ve\to 0}\z_\ve^E (\tau,G)=\lim_{\ve\to 0}\z^\ve
\bl(\ve \tau,G\br)
$$
is solution of the deterministic equation
$$
d \z^E(\tau)=E\z^E(\tau)d\tau, \Eq(0.2)
$$
where $E$ is the linearized Euler operator around the global
equilibrium. Equations \equ(0.2) are a system of linear hyperbolic
equations. The stochastic noise should appear as a correction of order
$\ve$ as
$$
d \z^E_\ve(\tau)=(A+\ve {\cal D})\z^E_\ve(\tau)d\tau+\sqrt\ve B dW_\tau
+O(\ve^2),
$$
where ${\cal D}$ is the linearized Navier-Stokes operator around the
global equilibrium and
$$
B B^*=-2\cd C
$$
is the fluctuation-dissipation relation.  Hence, to see a finite noise
one has to look at longer times $\tau=\ve^{-1}t$.  Formally, 
%change
%setting
since
$\z^\ve(t)=\z^E_\ve(\ve^{-1}t)$, we get
$$
d \z^\ve(t)=(\ve^{-1}E+ {\cal D})\z^\ve(t)dt+ B dW_t +O(\ve).
$$
Then the limit $\lim_{\ve\to 0}\z^\ve(t)$  does not exist because the
Euler modes are too big on this time scale. A similar difficulty is
present also in the case of ASEP but the analogous of $E$ is simply an
operator of the form $v\cdot\nabla_x$ with $v_i=(p_i-q_i)(1-2\a)$,
$p_i$, $q_i$ the rates of jumping to the left and right respectively
and $\a=\E[\eta]$.  Therefore, a Galilean shift is sufficient to remove
the divergence and in fact in [CLO] the central limit theorem is
proved for a fluctuation field of the form
$$
Y^\ve(t,G)=\ve^{d\over 2}\sum_x G(\ve x- \ve^{-1}vt)
\big[(\eta_{\ve^{-2}t}(x) -\E[\eta]\big].
$$
In our case a possible way to subtract the Euler modes is to consider a
modified fluctuation field which moves together with the waves
solutions of \equ(0.2), traveling with velocity of order
$\ve^{-1}$. 
%change
Denoting by $E^*$ the adjoint operator of $E$, 
we define the fluctuation field as
$$
\xi^\ve (t,G)=\z^\ve(t,e^{-{t\over\ve}E^*}G).
$$
We prove that the limit $\ve\to 0$ exists and satisfies a suitable
stochastic differential equation.  Before writing the equation, we
consider the same problem in a very simple case: let $A$ and $M$ be
$K\times K$ matrix with complex entries such that $A=-A^*$ where the
adjoint is relative to the scalar product in $\R^K$.  Consider the
linear ODE system
$$
\dot x_\ve=({\ve^{-1} A}+M) x_\ve ,\quad x_\ve(0)=\bar x.
$$
Then, $y_\ve=e^{-{t\over \ve}A}x_\ve$ is solution of
$$
\dot y_\ve=e^{-{t\over \ve}A}M e^{{t\over \ve}A}y_\ve, \quad
y_\ve(0)=\bar x.
$$
Consider the limit
$$
U:=\lim_{\ve\to 0}\int _0^1\! ds\, e^{-{s\over \ve}A}M e^{{s\over
    \ve}A}=\lim_{\ve\to 0}{\ve}\int_0^{1\over \ve}\!ds\, e^{-{s}A}M
e^{{s}A}
$$
An asymptotic average theorem [EP] states that for any $\d>0$ and
$T>0$ there exists $\ve_0>0$ such that the solution $z$ of
$$
\dot z= U z, \quad z(0)=\bar x
$$
satisfy
$$
\sup_{0<t<T}|y_\ve-z|<\d,\quad 0<\ve<\ve_0
$$
Therefore,
$y=\lim_{\ve\to 0}y_\ve$ is solution of
$$
\dot y=U y, \quad y(0)=\bar x.
$$
The limit $U $ can be characterized in the following way: Let
${\cal N}$ be the space of the $K\times K$ matrices with complex
entries.  ${\cal N}$ is a Hilbert space under the inner product
$$
(A,B)=\sum_{1\le i,j\le K}A^*_{ij}B_{ij}.$$
For $A\in {\cal N}$
define $\Pi_A$ as the orthogonal projection onto the subspace of the
matrices which commute with $A$
$$
\{B\in {\cal N}:[B,A]=0\}.
$$
Since the spectrum of $A$ is imaginary it is easy to see that
$$
U=\lim_{\ve\to 0}{1\over \ve}\int_0^{1\over \ve}\!ds\, e^{-{s}A}M
e^{{s}A}
$$
is the projector $\Pi_A M$.
  
\medskip

Applying this kind of considerations to our problem, by Fourier
analysis, we prove that $\xi (t,G)=\lim_{\ve\to 0}\xi^\ve (t,G)$
exists and is a stationary generalized Ornstein-Uhlenbeck process
characterized formally by the stochastic differential equation
$$
d \xi(t)= N\xi(t)dt+ B dW_t,
$$
with $ \quad BB^*=-2 N C$, where $C$ is the compressibility
matrix, $N$ is a second order differential operator and $BW_t$ is a mean
zero Gaussian field. In particular, this proves the
fluctuation-dissipation relation $-2NC=BB^*$ for this model.  Denoting
by $\hat M$ the Fourier transform of a $5\times 5$ matrix whose
entries are differential operators, we can characterize $N$ as
follows:
$$
\hat N= \Pi_{{\hat E}} \hat{\cal D},
$$
$\Pi_{{A}}$ the projection on the space of the operators commuting
with $A$. To conclude, we want to stress that this procedure of
subtracting the Euler modes works in this case because the equations
for the equilibrium fluctuations are linear.

\medskip
  
The central limit theorem for equilibrium fluctuations is a well
investigated topics [S], [KL]. A standard procedure is to establish
first the tightness of the sequence of fluctuation field. Then, the
study of the martingale problem allows to identify the unique weak
limit as a generalized Ornstein-Uhlenbeck process by the use of
Holley-Stroock theory. It is crucial to evaluate some expression in the
martingale problem in terms of the fluctuations
field.  This step, called Boltzmann-Gibbs principle, was first achieved
in [BR] for symmetric zero range process. The alternative method by [CY]
and  [C1] was extended to non gradient systems by [L] and [C2]. The
extension is based on a suitable modification of the fluctuation field by
adding lower order terms, determined by identifying the diffusion
coefficient in the hydrodynamic equations. In [CLO] this approach has been
extended to a non-symmetric case by proving a stronger tightness
result and as consequence a stronger Boltzmann-Gibbs theorem. We
extend the results on tightness and Boltzmann-Gibbs theorem in [CLO]
to the present model. Moreover, we prove the convergence of the time
averages of the form appearing in the martingale problem, by using and
adapting some results in [EP] which studied the convergence of
solutions to the linearized Navier-Stokes equations of solutions to
the linearized Boltzmann equation.  The paper is organized as follows.
In Section 2 we define the model and recall the previous results on
the hydrodynamic limit that we will need in the sequel. In Section 3
we define the fluctuation field and state the results. In Section 4 we
identify the limiting distribution of $Q^\ve$ by using Holley-Stroock
characterization of Ornstein-Uhlenbeck processes with martingales. The
Boltzmann-Gibbs principle is proved in Section 5 together with the
tightness of the process. The theorems stating the existence of the
time averages are in the Appendix.

\vskip 1truecm
\noindent{\bf 2. Model and hydrodynamic limit.}
\vskip .5truecm \numsec= 2 \numfor= 1 \numtheo=1

We consider the following model introduced in [BEM], which is a
generalization of the model in [EMY2]: given a finite set of velocities
$\cv\subset\R^3$, particles with velocity $v\in\cv$ evolve on the
sub-lattice $\L_L=\{-L,\ldots ,L\}^3$, with periodic boundary
conditions, according to an exclusion process. Collisions between two
particles can also occur provided that the momentum and the kinetic
energy are conserved. The set $\cv$ is chosen in the following way:
$$
\cv= \cv_1\cup\cv_2,
$$
where $\cv_1$ is made of  8 velocities given by
$$
\cv=\{(\pm1,\pm1,\pm 1)\} %\Eq(2.7)
$$
and $\cv_2$ contains  24 velocities, given up to permutation by
$$
(\pm\vp,\pm1,\pm 1), \Eq(2.8)
$$
where $\vp$ is some irrational number suitably chosen.

Formally, if we denote by $\n(x,v)\in\{0,1\}$ the number of particles
on site $x\in\L_L$ with velocity $v\in\cv$, then the infinitesimal
generator of the dynamics on the space $\Omega_L=\{\n=(\n(x,v),\,
x\in\L_L,\, v\in\cv)\}$ is defined as
$$
\cl=\cle+\clc, %\Eq(o2.1)
$$
where $\cle$ is the generator of the nearest neighbor exclusion
process with different colors (velocities) and $\clc$ the generator of
the collision process. For a local function $f$ on $\Omega_L$, $\cle$
is given by
$$
\cle f(\n)=\sum_{v\in\cv}\sum_{|e|=1}\sum_{x\in\L_L}
\bigr(\chi+{1\over 2}e\cdot v\bigr)\eta(x,v)
\left[f\bigl(\n^{x,x+e,v}\bigr)-f\bigl(\n\bigr)\right], %\Eq(o2.2)
$$
where $e$ is a unitary vector of $\Z^3$ ($e_\a$, $\a=1,2,3$, will
denote the unitary vectors with positive coordinates),
$\n^{x,x+e,v}$ is the configuration obtained after exchanging
the values of $\n(x,v)$ and $\n(x+e,v)$ and $\chi$ is some positive
constant large enough such that the jump rate is positive. Note that it
is chosen so that the drift of the particles with velocity $v$ is exactly
$v$. 

The collisions generator 
$\clc$ is given by
$$
\clc f(\n)=\sum_{x\in\L_L}\sum_{q\in\cq}
\left[f\bigl(\n^{x,q}\bigr)-f\bigl(\n\bigr)\right], %\Eq(o2.3)
$$
where $\cq$ is the set of admissible collisions, namely the set of
velocity quadruples $q=(v,w,v',w')\in\cv^4$ such that $v+w=v'+w'$ and
$|v|^2+|w|^2=|v'|^2+|w'|^2$, and $\n^{x,q}$ is the configuration
obtained after the collision on site $x$ between two particles with
incoming velocities $v,w$ and outgoing velocities $v',w'$. Notice that
in order to preserve the exclusion rule, we take $\n^{x,q}$ unchanged
with respect to $\n$ if one of the conditions $\n(x,v)=0$,
$\n(x,w)=0$, $\n(x,v')=1$ or $\n(x,w')=1$ is fulfilled.

We denote by $\eta_x=\{\eta(x,v),\, v\in {\cal V}\}$ the particle
configuration in $x\in \L_L$.
For a configuration $\n$, the mass, momentum and kinetic energy in
site $x$ are
$$
%\eqalignno{
\eqalign{ I_0(\n_x) & =\sum_{v\in\cv}\n(x,v), %& \eq(o2.4)
  \cr
  I_\a(\n_x) & =\sum_{v\in\cv}(v\cdot e_\a)\n(x,v)\,,\ \a=1,2,3, %&\eq(o2.5)
  \cr I_{4}(\n_x) & =\sum_{v\in\cv}{1\over 2}|v|^2\n(x,v). 
  %& \eq(o2.6)
  \cr}
$$
It is easy to check that the quantities
$N_\b(\n)=\sum_xI_\b(\n_x)$, $\b=0,\ldots,4$, are conserved by the
full dynamics. It is shown in [BEM] that, by choosing suitably the
parameter $\vp$ in \equ(2.8), they are the only conserved quantities,
in other words this model has the property of local ergodicity. 

As a consequence, the grand canonical measures below are invariant
for $\cl$
$$
\mu_{L,n}(\n)=Z_{L,n}^{-1}\prod_{x\in\L_L}\exp
\Bigl\{\sum_{\b=0}^{d+1}n_\b I_\b(\n_x)\Bigr\}, \Eq(o2.9)
$$
where $n=(n_0,\ldots,n_4)\in\R^5$ are the chemical potentials
and $Z_{L,n}$ is a normalization
constant.  All these product measures are absolutely continuous with
respect to the reference measure $\mu$ obtained by taking
$n$ as $n_0:=(r,0,0,0,\theta)$.  We set $m_\b=\E^\mu[I_\b(\n_0)]$ 
for $\b=0,\ldots,4$
(notice that $m_\b=0$ if $\b=1,2,3$) and $\tilde I_\b=I_\b-m_\b$.

\medskip

In the sequel we call $\ve=L^{-1}$.
The law of the process $(\n_t(x,v))$ with
generator $\ve^{-2}\cl$ starting from $\mu$ is denoted by $\P_\ve^\mu$
and the corresponding expectation by $\E_\ve^\mu$.
We also call $f_0(v)=\E^{\mu}[\n(x,v)]$ the density of particles with
velocity $v\in\cv$ with respect to the reference measure $\mu$. For any
function $h$ on $\cv$, we put
$\ps{h}=\sum_{v\in\cv}h(v)$. 

The currents $w_{x,\a}^{\b}$ of the conserved quantities 
$I_\b$, $\b=0,\ldots,4$, 
at site $x$ in direction $e_\a$, $\a=1,2,3$, are defined by
$$
\cl I_\b(\n_x)=\sum_{\a=1}^3\nabla_\a^{-}w_{x,\a}^{\b}, %\Eq(3.4)
$$
where, if $g$ is a function on $\Lambda_L$,
$$
\na^-_\a g(x)=(\na_\a g)(x-e_\a)\quad {\rm and}\quad \na_\a
g(x)=g(x+e_\a)-g(x). %\Eq(3.5)
$$
Since the local quantities $I_\b(\n_x)$ are conserved by the
collision generator, we have $\cl I_\b(\n_x)=\cle I_\b(\n_x)$ and the
currents can be written as the sum of a symmetric and an antisymmetric
parts
$$
w^\b_{x,\a}=\chi\na_\a I_\b(\n_x)+w^{(a),\b}_{x,\a} %\Eq(3.6)
$$
and
$$
\eqalign{ w^{(a),0}_{x,\a} &= \ps{v_\a b_{x,\a}(v)},\quad
  w^{(a),\b}_{x,\a} = \ps{ v_\a v_\b b_{x,\a}(v)},\  \b=1,2,3\cr
  w^{(a),4}_{x,\a} &= {1\over 2}\ps{v_\a|v|^2b_{x,\a}(v)},\cr
  } %\Eq(3.5a)
$$
with
$$
b_{x,\a}(v)=\n(x+e_\a,v)\n(x,v)-{1\over
  2}\bigl(\n(x+e_\a)+\n(x,v)\bigr). %\Eq(3.5b)
$$

\bigskip

Let ${\cal G}$ be the space of local functions $h$ on $\Lambda_L$ such
that
$$
\E^\mu[h]=0\quad \hbox{\rm and}\quad {\p\E^{\mu_{L,n}}[h]\over
  \p m_\b(n)}{\big|_{n=n_0}}=0,\quad \b=0,\ldots,4,
\Eq(2.28)
$$
where $m_\b(n)=\E^{\mu_{L,n}}[I_\b]$. 
In view of the application of the
Boltzmann-Gibbs principle, it is important to modify the currents
$w^{(a),\b}_{x,\a}$ so that they are in the space ${\cal G}$. It is
enough to subtract suitable combinations of the conserved quantities and
we now get their explicit expressions. 

Let $n$ be the chemical potential $n=n_0+\d
n=(r+\d n_0,\d n_1,\d n_2,\d n_3,\theta+\d n_4)$, then
$$
\eqalign{ \E^{\mu_{L,n}}\bigl[w^{(a),0}_{0,\a}\bigr] -c_\a^0 &=
  {1\over 3}\ps{|v|^2h_1}\d n_\a+o(\d n),\cr
  \E^{\mu_{L,n}}\bigl[w^{(a),\b}_{0,\a}\bigr] -c_\a^\b &=
  \d_{\a,\b}\Bigl[{1\over 3}\ps{|v|^2h_1}\d n_0+{1\over
    6}\ps{|v|^4}h_1\d n_0\Bigr]\d n_\a+o(\d n),\ \quad\b=1,2,3,\cr
  \E^{\mu_{L,n}}\bigl[w^{(a),4}_{0,\a}\bigr] -c_\a^4 &= {1\over
    6}\ps{|v|^4h_1}\d n_\a+o(\d n),\cr } 
%\Eq(3.7)
$$
where $h_0=f_0(1-f_0)$, $h_1=h_0(1-2f_0)$  and
$$
c_\a^\b=\E^\mu\bigl[w^{(a),\b}_{0,\a}\bigr]
\Eq(3.7bis)$$
If we denote by $\d m_\b= \E^{\mu_{L,n}}[I_\b(\n_0)]-
m_\b$, we get
$$
\eqalign{ \d n_0 = & {1\over \Phi}\bigl(\ps{|v|^4h_0}\d m_0
  -2\ps{|v|^2h_0}\d m_{4}\bigr),\cr \d n_\a = & {3\over
    \ps{|v|^2h_0}}\d m_\a,\cr \d n_{4} = & {2\over
    \Phi}\Bigl(2\ps{h_0}\d m_{4}- \ps{|v|^2h_0}\d m_0\Bigr),\cr }
%\Eq(3.7a)
$$
where
$$
\Phi=\ps{|v|^4h_0}\ps{h_0}-\ps{|v|^2h_0}^2>0. %\Eq(3.7b)
$$
So, defining for $\b,\nu=0,\ldots,4$ and $\a=1,2,3$,
$$
d_\a^{\b,\nu}={\p\E^{\mu_{L,n}}\bigl[w^{(a),\b}_{x,\a}\bigr]\over \p
  m_\nu(n)}{\big|_{n=n_0},} %\Eq(3.7c)
$$
we obtain
$$
d_\a^{\b,\nu}=b_0\d_{\b,0}\d_{\a,\nu} +b_4\d_{\b,4}\d_{\a,\nu}+
\1_{\{1,2,3\}}(\b)\d_{\a,\b}\bigl[a_0\d_{\nu,0}+a_4\d_{\nu,4}\bigr],
\Eq(3.7d)
$$
with
$$
b_0 ={\Phi_2\over 3\Phi},\qquad b_4 =2{\Phi_1\over 3\Phi},\qquad a_0
={\ps{|v|^2h_1}\over\ps{|v|^2h_0}},\qquad a_4
={\ps{|v|^4h_1}\over\ps{|v|^2h_0}}, \Eq(3.7e)
$$
   
$$
\eqalign{ \Phi_1
  &=\ps{h_1|v|^4}\ps{h_0}-\ps{h_1|v|^2}{\ps{h_0|v|^2}},\cr \Phi_2
  &=\ps{h_0|v|^4}\ps{h_1|v|^2}-\ps{h_1|v|^4} \ps{h_0|v|^2}.\cr }
%\Eq(2.23)
$$
Therefore the local function
$$
g^\b_{\a}(\n)=
w^{(a),\b}_{0,\a}-c^\b_\a-{1\over 2}\sum_{\nu=0}^4d_\a^{\b,\nu}\bl(\tilde
I_\nu(\n(0))+\tilde I_\nu(\n({e_\a}))\br) \Eq(3.13a)
$$
belongs to $\cg$.

\medskip

\bigskip
\noindent{\it Slow-fast modes decomposition of the currents}

We denote by $\ov I^+_\ell=(\ov I_{0,\ell},\ldots,\ov I_{4,\ell})$ the
empirical averages of the conserved quantities over the block
$\L_\ell$ of length $\ell$:
$$
\ov I_{\b,\ell}={1\over (2\ell
  +1)^3}\sum_{|y|\le\ell}I_\b(\n_y),\quad \b=0,\ldots,4. %\Eq(2.70)
$$
The measure $\hat\mu_{\ell,m}$, $m\in\R^5$ is defined as the canonical
Gibbs state of $(2\ell+1)^3$ sites with parameters such that $\ov
I^+_\ell=m$. It is the uniform probability on the set $\O_{\ell,m}$ of
configurations on the block $\L_\ell$ such that $\ov I^+_\ell=m$.  We
denote by $\a_\ell(g)$ the conditional expectation of $g$ given the averages
$\ov I^+_\ell$
$$
\a_\ell(g)=\E^{\mu}[g|\ov I^+_\ell]. %\Eq(2.71)
$$

We call $\cl_{s,\ell}$ the symmetric part of the generator $\cl$
restricted to the block $\L_\ell$. Since the measures $\hat\mu_{\ell,m}$ are the
only extremal invariant measures for $\cl_{s,\ell}$, we can
define $\cl^{-1}_{s,\ell}g$ for any function $g$ such that
$\a_\ell(g)=0$. Given any local function $g$ on $\O_\ell$, the finite
volume ``variance'' $V_\ell(g,n)$ is
$$
V_\ell(g,n)={1\over (2\ell_1+1)^3}\E^{\mu_{\ell,n}}\Bigl[
\Bigl(\sum_{|x|\le\ell_1}
  (\tau_xg
  -\a_\ell(g))\Bigr)(-\cl_{s,\ell})^{-1}\Bigl(\sum_{|x|\le\ell_1}
  (\tau_xg -\a_\ell(g))\Bigr)\Bigr], %\Eq(2.72)
$$
where $\tau_\cdot$ is the translation operator on $\O_L$,
$\tau_xg(\n)=g(\tau_x\n)$, 
$\ell_1=\ell-\ell^{1/9}$, $\ell$ large enough. The
``variance'' $V(G,n)$ of $G$ is given by
$$
V(G,n)=\limsup_{\ell\to\infty}V_\ell(G,n). \Eq(o2.73)
$$
With an abuse of notation, we denote $V_\ell(G,n)$ by
$V_\ell(G,r,\theta)$ when $n$ is the chemical potential
$n_0=(r,0,0,0,\theta)$.

\medskip

% For an integer $\ell$, we let $\ov\ell=\ell^{5}$ and
% $k=\ell\ve^{-2/3}$. We assume that the box $\L_k$ is divided into
% cubes $\L_{\ell,\s}$ of size $(2\ov\ell+1)$ with centers
% $\s\in(2\ov\ell+1)\Z^3$, $|\s|\le k$. We consider the sub-cubes
% $\L_{\ov\ell_1,\s}$ of size $\ov\ell_1=\ov\ell-\ell^{1/9}$ and their
% union is denoted by $\hat\L_k=\cup_{|\s|\le k}\L_{\ell_1,\s}$. The
% functions $\o$ and $\hat\o$ are the normalized indicator functions
% $$
% \o(x)=(2k+1)^{-3}\1\{x\in\L_k\},\quad
% \hat\o(x)=|\hat\L_k|^{-1}\1\{x\in\hat\L_k\} \Eq(3.5.c)
% $$

% \medskip
   
\goodbreak
We state here 
 the  results in [BEM] 
\bigskip

\noindent{\bf \Theorem (so3.1b).}
{\it There exists a %positive diffusion matrix
rank 2 tensor $\bar D=\bl(\bar
  D^{\b,\nu}_{\a,\g}\br)$ ($\bar D^{\b,\nu}$   positive definite matrix)
and a sequence of local functions
$h^{(q)}=(h_\a^{(q),\b},\,{\a=1,2,3,\,\b=0,\dots,4})$ 
in ${\cal G}$ such that, setting
$$u^{(q),\b}_{\a}(\n)=g^\b_{\a}(\n) -\sum_{\g=1}^3\sum_{\nu=0}^4\bar
D^{\b,\nu}_{\a,\g}\nabla_\g \tilde I_\nu(\n(0)) -\cl h^{(q),\b}_\a(\n),
\Eq(3.14)
$$
where $g^\b_{\a}(\n)$ is defined in \equ(3.13a), it results
$$
\limsup_{q\to\infty}\sum_{\a=1}^3\sum_{\b=0}^4V(u_\a^{(q),\b},r,\theta)=0.
%\Eq(3.8d)
$$}
\smallskip
\noindent Above Theorem actually holds for any function in ${\cal G}$.
\smallskip
\noindent{\bf \Lemma (so4.6).}  
{\it The tensor $\bar D$ satisfies
  $$
  a\cdot(\bar D C) a=\lim_{q\to \infty}\E^\mu\bl[\Gamma(a\cdot h^{(q)})
  (-\cl_s)(a\cdot h^{(q)})\br].
  \Eq(4.15)
  $$
}
In this formula,
$a\cdot b=\sum_{\a=1}^3\sum_{\b=0}^4a_\a^\b b_\a^\b$, $\G
(g)=\sum_x\tau_x g$, $\cl_s$ is the symmetric part of $\cl$ in
$L^2(\mu)$, $C$ is  the $5\times5$ compressibility
matrix (see \equ(comp) below for an explicit expression) and $\bar DC$ is the
tensor $(\bar DC)^{\b,\nu}_{\a,\g}=(\bar D_{\a,\g}C)^{\b,\nu}$. 
We define $D=\bar D+\chi\I$ where 
$\I^{\b,\nu}_{\a,\g}=\d_{\a,\g}\d_{\b,\nu}$.

\bigskip

%We also define the temperature current as
%   $$
%\bar w^\b_{x,\a}=\chi\na_\a I'_4(\n_x)+\bar w^{(a),4}_{x,\a}
%   \Eq(3.7)
%   $$
%and for notational convenience, we let $\bar w^{(a),\b}_{x,\a} =
%w^{(a),\b}_{x,\a}$ for $\a=1,2,3$ 

\bigskip
\noindent{\it Hydrodynamic limit}

Given functions $n_\b(x)$, $\b=0,\ldots ,4$, we consider the Gibbs
states with chemical potential $n(x)=(n_0(x),\ldots,n_4(x))$
$$
\mu_{L,n}(\n)=Z_{L,n}^{-1}\prod_{x\in\L_L}\exp
\Bigl\{\sum_{\b=0}^{4}n_\b(x) I_\b(\n_x)\Bigr\}. %\Eq(o2.10)
$$

Now, assume that the initial distribution of the particles is
$\mu_{L,n}$ with $n=(n_\b)$ the slowly varying chemical potentials
given by
$$
n_\b(x)=\l_\b^{(0)}+\ve\l_\b^{(1)}(\ve x)+\ve^2\l_\b^{(2)}(\ve x),
\Eq(o2.11)
$$
where $\l^{(0)}=(\l_\b^{(0)})=n_0$ and $\l_\b^{i}$ are
smooth functions on the 3-d torus $\T_3$. 
We define the local equilibrium measure as the Gibbs
states $\mu_{L,n(\cdot,t)}$ with $n(\cdot,t)$ the chemical potential
given by
$$
n_\b(x,t)=\l_\b^{(0)}+\ve\l_\b^{(1)}(\ve x,t)+\ve^2\l_\b^{(2)}(\ve
x,t). \Eq(o2.12)
$$
Furthermore, we assume
$$
\div\, \underline{\lambda}^{(1)}=0,\quad <h_1v^2>\l_0^{(1)}+{1\over
  2}<h_1v^4>\l_4^{(1)}=0.
$$

Then in [BEM] (see also [EMY2]) it has been proved that the the law of
the process at time $t>0$ is well approximated by the local
equilibrium in the sense that the relative entropy per unit volume of
the non-equilibrium measure with respect the local equilibrium times
$\ve^{-2}$ vanishes in the limit $\ve\to 0$.

\medskip

We can now state the result proved in [BEM] on the
hydrodynamic limit.  Let $u(z,t)$, $z\in\T_3$, $t\in[0,t_0]$,
$t_0>0$, be the classical smooth solutions of the following
Navier-Stokes equation
$$
\eqalign{ & \div\, u =0, \cr & \p_t u_\b+\partial_\b p+Ku\cdot\na u_\b
  = \sum_{\a=1}^3 D_{\a,\b}\p_\a^2 u_\b,\quad \b=1,2,3,\cr } \Eq(2.21)
$$
with initial condition
$u_\a(z)=\E^{\mu_{L,n(\ve^{-1}z)}}[I_\a(\n_0)]$ and let $\ce(z,t)$ be
the solution of the energy equation
$$
{\p\over\p t}\ce +Hu\cdot\na \ce =\sum_{\a=1}^3 {\cal
  K}_\a(\p_\a^2\ce), \Eq(2.21a)
$$
with initial condition
$\ce(z)=\E^{\mu_{L,n(\ve^{-1}z)}}[I_4(\n_0)]$.  The constants appearing
in \equ(2.21) and \equ(2.21a) are given by 
$$
K=18{\ps{v_1^2v_2^2 h_2}\over \ps{h_0|v|^2}^2},$$
with
$h_2={1\over 2}h_1(1-6f_0(1-f_0))$
and
$$\quad H= {1\over
  \ps{h_0|v|^2}}{\Psi_1-2C\Psi_2\over \Phi_2+C\Phi_1},\quad C ={1\over
  2}{\ps{h_1|v|^4}\over \ps{h_1 |v|^2}}, %\Eq(2.22)
$$
where
$$
\eqalign{ \Psi_1&=\ps{ h_2|v|^6}\ps{h_1|v|^2}-\ps{
    h_2|v|^4}\ps{ h_1|v|^4},\cr 
  \Psi_2&=\ps{ h_2|v|^4}\ps{h_1|v|^2}-
    \ps{ h_2|v|^2}\ps{ h_1|v|^4}.\cr 
}
$$
  
\bigskip
\goodbreak

Let $\P^{\mu_{L,n}}_\ve$ be the law of the process $\n_t(x,v)$ with
generator $\ve^{-2}\cl$ starting from the measure $\mu_{L,n}$ defined in
\equ(o2.11), with chemical potentials $n_\a(x)$ of the form
\equ(o2.12). The density ($\nu_{0}^\ve(t,z)$), the momentum
($(\nu_{\b}^\ve(t,z))_{\b=1,2,3}$) and energy ($\nu_{4}^\ve(t,z)$) 
empirical fields are defined as
$$
  \nu_{\b}^\ve(z,t) =\ve^2\sum_{x\in\L_L} \d(z-\ve
    x)\tilde I_{\b}(\n_t(x)),
%\Eq(2.24)
$$
where $\tilde I_\b(\n_x)=I_\b(\n_x)-m_\b$,
$m_\b=\E^{\mu}[I_\b(\n_0)]$ and 
$\n_t(x)=\{\n_t(x,v), \, v\in {\cal V}\}$.    

\bigskip

\noindent{\bf \Theorem (s2.1)}
\nobreak
{\it 
  The density, momentum
  and energy  empirical fields converge, for 
  $t\le t_0$, weakly (in space) in\/ $\P^{\mu_{L,n}}_\ve\!\!$ probability,
  to $\rho(z,t)dz$, $u(z,t)dz$ and
  $\ce(z,t)dz$, where $a\rho+b\ce=c$ for suitable $a,b,c$.}
\smallskip
Note that the transport coefficients $D_{\a,\b}$ and ${\cal
 K}_\a$ are suitable combinations of the diffusion coefficients 
 $D^{\b,\nu}_{\a,\g}$ in Theorem \equ(so3.1b). The explicit expressions are given in [BEM],
but we omit them because they do not play any role in this paper.
\bigskip

\vskip 1truecm \goodbreak
\noindent{\bf 3. Fluctuation field and results}.
\vskip .5truecm \numsec= 3 \numfor= 1 \numtheo=1 

In this paper, we are interested in the equilibrium fluctuations of
the mass, momentum and energy fields. The initial fluctuations,
distributed in terms of the measure $\mu$, are finite but they may become
infinite at later very long times because of the effect of waves
moving with velocity $\ve^{-1}$, which are the solutions of the
linearized (around the equilibrium) Euler equations (linear hyperbolic
equations) for this model. To remove the diverging terms we have to
modify the usual definition of fluctuation fields not simply by a
shift but considering fluctuations which move together with the
traveling waves

We denote by $U^\ve_t$ the operator $\exp(-{t\over\ve} E^*)$ where $E$
is the linearized Euler operator, a $5 \times 5$ matrix whose entries
are first order differential operators with constant coefficients,
%change
$$
E=\pmatrix{0&-a_0\underline{\partial}&0\cr
-b_0{\underline{\partial}}&0&-b_4{\underline{\partial}}\cr
  0&-a_4\underline{\partial}&0\cr} 
%\Eq(E)
$$
and $*$ is the adjoint with respect the
usual scalar product in $L^2(\T_3,\R^5)$ 
(the constants $a_i$ and $b_i$ are defined in
\equ(3.7e) and ${\underline{\partial}}=(\partial_1,\partial_2,\partial_3)$ is the
gradient operator).

\bigskip

For any smooth function $G=(G_\b)_{\b=0,\ldots,4}:\T_3\rightarrow
\R^5$ consider the (scalar) fluctuation field $\xi^\ve$ on the state space 
$(\T_3)^{\otimes 5}$
$$
\xi^\ve\bl(t,G\br)=\ve^{3/2}\sum_{\b=0}^4\sum_x (U^\ve_t
G)_{\b}(\ve x)\tilde I_\b(\n_t(x)). \Eq(2.26)
$$

It is equivalent to consider
the vector fluctuation field $(\xi^\ve_\b)_{\b=0,\ldots 5}$ on $\T^3$ 
whose components $\xi^\ve_0$,  $(\xi^\ve_\b)_{\b=1,\ldots 3}$ and
$\xi^\ve_4$ are respectively the
density, momentum and energy fluctuation fields, defined as
$$
\xi^\ve_\b(t,\varphi)=\xi^\ve(t,G^{(\b)}),\quad \b=0,\dots,4,
$$
where $G^{(\b)}$ is the vector function with only the $\b$
component non vanishing and $G^{(\b)}_\b=\varphi$.
  
%  -E^*=\left(\matrix{0&b_0{\bf \nabla}&0\cr
%  a_0{\bf \nabla}&0&a_4{\bf \nabla}\cr
%  0&b_4{\bf \nabla}&0\cr}\right)
%  $$
  
\bigskip

We want to study the evolution of the fluctuation fields in the limit
$\ve\to 0$ when the fields are initially distributed with the
equilibrium measure $\mu$, given by \equ(o2.9). 
% We omit
%the dependence on the chemical parameters $n$. 
We notice that the
initial covariance of the limiting fields $\lim_{\ve\to 0}\xi^\ve
_\b(0,\varphi)=\xi_\b(0,\varphi)$ is
$$\E^\mu_\ve\Bigl[ \xi_\b(0,\varphi)\xi_\nu(0,\psi)\Bigr]=C_{\b,\nu}
\int_{\T_3}\!dx\, \varphi(x)\psi(x),
\Eq(cov)$$
where $C$ is the compressibility matrix ($5 \times 5$)
$$
C=\pmatrix{\ps{h_0}&0&\ps{h_0{|v|^2\over 2}}\cr 0&{1\over
    3}\ps{|v|^2h_0}\Bbb I_3&0\cr \ps{h_0{|v|^2\over 2}}
&0&\ps{h_0{|v|^4\over 4}}\cr},
\Eq(comp)$$
with  $\Bbb I_3$ the $3\times 3$ identity matrix, $h_0$  defined 
in the paragraph before \equ(3.7bis) and
$\ps{\,\cdot\,}$ in the paragraph after
\equ(o2.9).
\smallskip

Remark that $E$ is not anti-hermitian in $L^2(\T_3,\R^5)$, since $a_0\ne b_0$ and
$a_4\ne b_4$. However a
straightforward computation shows that $EC$ satisfies $EC+CE^*=0$.

We want to show that the fluctuation field converges to a stationary
Gaussian vector process with a given covariance. The equal time
covariance is exactly \equ(cov) because of the stationarity of the
limiting process.

\bigskip
            
To state the results we need some extra notation.
%If $S_{-k}$
%is the Sobolev space of order $-k$ on $\T^3$, then we also consider
%the Hilbert space $H^C_{-k}$ endowed with the inner product
We introduce the Hilbert spaces ${\cal H}_{k}, k\in\Z$ defined by the
scalar product
$$\ps{G,H}_{k}=\ps{G,L^kH}_0,$$
where
$L=I-\Delta$, $\Delta$ the Laplacian operator
and $\ps{.,.}_0$ is the usual inner product of $L^2(\T_3,\R^5)$:
$$
\ps{G,H}_0=\sum_{\b=0}^4\int_{\T_3}\!dx\,G_\b(x)H_\b(x).
\Eq(fps)
$$
Denote by $\|\cdot\|_k$ the norm of
$\ch_{k}$ and by ${\cal H}_{-k}$ the dual of ${\cal H}_{k}$ with respect to
the inner product of $L^2(\T_3,\R^5)$.
The fluctuation field $(\xi^\ve(t))_{t\ge 0}$ is a distribution valued
stochastic process taking
values in the Sobolev space ${\cal H}_{-k_0}$ for some suitable $k_0$.
Its path space is $D([0,T], {\cal H}_{-k_0})$ ($T>0$), the space of functions
with values in ${\cal H}_{-k_0}$, right continuous with left limits,
endowed with the uniform (in time)  weak (in space) topology.  
We call $Q^\ve$ the law of
$(\xi^\ve(t))_{t\ge 0}$ when the process is initially distributed
according to the equilibrium measure $\mu$. It is therefore a
probability measure on the space $D([0,T],{\cal H}_{-k_0})$. 
%We denote
%by $P^\ve_\mu$ the law of the process $\eta_{\ve^{-2}t}$ with initial
%distribution $\mu$.

By analogy with \equ(fps),
we define for local functions
$g=(g_0,\ldots,g_4)$ on $\O_L$ %with vanishing equilibrium expectation
and smooth functions
$G=(G_0,\ldots,G_4)$ on $\T_3$
$$
\ps{G,g}_{0,L}=\ve^{3/2}\sum_{\b=0}^4\sum_{x\in\L_L}G_\b(\ve x)\tau_xg_\b.
\Eq(fps2)
$$
So the fluctuation field \equ(2.26) can be rewritten as
$$
\xi^\ve(t,G)=\ps{U^\ve_tG,\tilde I(\n_t(0))}_{0,L}.
$$
\vskip.5cm
We introduce the linearized Navier-Stokes 
operator ${\cal D}$ as
$$
{\cal D}G=\sum_{\a,\g=1}^3
D_{\a,\g}\pt_\a\pt_\g G.
\Eq(D)
$$
Then $N=\pi_E({\cal D})$ is the operator 
defined as the limit
$$
\lim_{\ve\to 0}{1\over T}\int_0^T\!dt\,
\exp\bl(-{t\over\ve } E\br) {\cal D}\exp \bl({t\over\ve
  }E\br)G=\pi_E({\cal D}) G
$$
and $\pi_E({\cal D})^*$ the adjoint of $N$ with respect to the inner product
$\langle\cdot,\cdot\rangle_0$ in $L^2(\T_3,\R^5)$.
The main result of this paper is
\bigskip

\noindent{\bf \Theorem (so2.2)}
{\it The probability measures $(Q^\ve)$ converge weakly in
  $D([0,T],{\cal H}_{-k_0})$ to the law $Q$ of the stationary generalized
  Ornstein-Uhlenbeck process $\xi$ with mean $0$ and covariance
  $$
  \E^Q\Bigl[\xi_\b(s,\phi)\xi_\nu(t,\psi)\Bigr]=
  \int_{\T_3}\!dx\,
  \bl((CS_{|t-s|})_{\b,\nu}\phi\br)(x)\psi(x),
  $$
  where $(S_t)_{t\ge 0}$ is the
  semi-group in $L^2(\T_3,\R^5)$ associated to $\pi_E({\cal D})^*$ and $C$ the
  compressibility matrix.
  It is formally characterized by the SDE
  $$\eqalign{
    d \xi(t) &= N\xi(t)dt+ B dW_t,\cr
    BB^* &=-2 NC.
    }
  $$
}

One of the main ingredient needed while studying the equilibrium
fluctuations is the so-called Boltzmann-Gibbs principle which states
that the non conserved quantities arising in the conservation laws may
be replaced by linear combinations of the conserved ones. In the
context of a non gradient system, the usual statement is not valid and
some corrections to the fluctuation field have to be introduced (see
[C], [Lu]). The situation in the case of an asymmetric system is more
delicate since the usual Boltzmann-Gibbs estimate is not sharp enough
and one has to prove a stronger result ([CLO]). We need to generalize 
such a result to the present setup. Indeed we prove the following

\bigskip
\goodbreak

%change
\noindent{\bf \Theorem (s2.3) (Boltzmann-Gibbs principle)}
{\it Assume that $h\in{\cal G}$ (see \equ(2.28)).  
Then, for any smooth function
 $G:\R_+\times\T_3\to \R $, the following estimate holds
  $$
  \limsup_{\ve\to 0}\E^\mu_\ve\left[\sup_{0\le t\le
      T}\left(\ve^{3/2-1}\int_0^t\!\sum_xG(s,\ve x)\tau_xh(\n, s)\,
      ds\right)^2\right]\le c\, T\|G
  \|_0^2\,V(h;r,\theta), %\Eq(2.29)
  $$
  \nobreak
  where $V$ is the infinite volume variance defined
  in \equ(o2.73).  }

\vskip1truecm

\noindent{\bf 4. Limiting distribution of the fluctuation field.}
\numsec= 4 \numfor= 1 \numtheo=1
\bigskip

The theory of Holley-Stroock [HS] characterizes the law $Q$ of the 
Ornstein-Uhlenbeck process $\xi$ described in Theorem \equ(so2.2) 
by the following martingale problem:
$$
\eqalign{ M_1(t,G) &
  =\xi(t,G)-\xi(0,G)-\int_0^t\!ds\,\xi(s,\pi_E({\cal D})^*G),
  \cr M_2(t,G) &
  =\bigl(M_1(t,G)\bigr)^2+2t\,\ps{G,C\,\pi_E({\cal D})^*G}_0\cr }
\Eq(mpb)
$$
are martingales under $Q$. In this section, we will prove that any limit law
$\bar Q$ of $Q^\ve$ satisfies \equ(mpb). Therefore from the
tightness of $(Q^\ve)$ (see Theorem \equ(tight) of section 5), 
it has to converge to $Q$ and Theorem \equ(so2.2) follows. 

The  processes analogous to \equ(mpb) for $Q^\ve$ are
$$
\eqalign{ M^\ve_1(t,G) &
  =\xi^\ve(t,G)-\xi^\ve(0,G)-\int_0^t\!ds\,\xi^\ve(s,\pi_E({\cal D})^*G),
  \cr M^\ve_2(t,G) &
  =\bigl(M^\ve_1(t,G)\bigr)^2+2t\,\ps{G,C\,\pi_E({\cal D})^*G}_0\cr }
%\Eq(mpb2)
$$
and we want to show that these processes are martingales up to some
error terms which vanish as $\ve$ goes to 0.
Given local functions $h=(h_\a)_{\a=1,2,3}=
(h^{\b}_\a)_{\a=1,2,3;\,\b=0,\ldots
 ,4}\in\cg$, we introduce the modified
%  vector fluctuation field as
%$$
% \z^\ve_\b(t,G,h)=\xi^\ve_\b(t,G) - \ve^{3/2+1}\sum_{\a=1}^3
% \sum_{\nu=0}^4\sum_x\p_\a (U^\ve_t G^{(\b)})_{\nu}(\ve
% x,\ve^{-1}t)\tau_xh^\nu_\a \Eq(3.1)
% $$
% and the scalar 
fluctuation field
% $$\eqalign{ \z^\ve(t,G,h)&=\ve^{3/2}\sum_{\b=0}^4\sum_x
%   (U^\ve_t G)_{\b}(\ve x)\tilde I_\b(\n_x)\cr &\quad -
%   \ve^{3/2+1}\sum_{\a=1}^3 \sum_{\b=0}^4\sum_x\p_\a (U^\ve_t
%   G)_{\b}(\ve x)\tau_xh^{\b}_\a }
% $$
$$
\z^\ve(t,G,h)=\xi^\ve(t,G)-\ve\sum_{\a=1}^3
  \ps{\p_\a(U^\ve_tG),h_\a}_{0,L},
$$
where $\ps{\cdot,\cdot}_{0,L}$ was defined in \equ(fps2). 
%change
Actually we will choose for $h$ the terms of the sequence $h^{(q)}$
defined in Theorem \equ(so3.1b), but we will omit the label $q$ for sake of
shortness.
It is clear that the
difference between $\z^\ve(t,G,h)$ and $\xi^\ve(t,G)$
vanishes in $L^2(\P^\mu_\ve)$ with $\ve$. Moreover, it is well known
that the following processes are martingales with
respect to the usual filtration related to the process $(\n_t(x,v))$
$$
\eqalign{ M^\ve_1(t,G,h) &
  =\z^\ve(t,G,h)-\z^\ve(0,G,h)-\int_0^t\g^\ve_1(s,G,h)\,
  ds,\cr M^\ve_2(t,G,h) &
  =\bigl(M^\ve_1(t,G,h)\bigr)^2-\int_0^t\g^\ve_2(s,G,h)\, ds,\cr }
\Eq(3.2)
$$
with
$$
\eqalign{ \g^\ve_1(t,G,h)
  &=\bigl(\p_t+\ve^{-2}\cl\bigr)\z^\ve(t,h,G),\cr
  \g^\ve_2(t,G,h) & =
  \bigl(\p_t+\ve^{-2}\cl\bigr)\bigl(\z^\ve(t,h,G)^2\bigr)
  -2\big(\z^\ve(t,h),G\big)
  \bigl(\p_t+\ve^{-2}\cl\bigr)\big(\z^\ve(t,h),G\big).\cr } 
%\Eq(3.3)
$$

\bigskip

We first compute the compensator $\g^\ve_1$. Let
$w^{(a)}_{x,\a}=(w^{(a),\b}_{x,\a})_{\b=0,\ldots ,4}$. Then
$$
\eqalign{ \g^\ve_1(t,G,h) = &
  \ps{\p_t(U^\ve_t G),\tilde I(\n_t(0))}_{0,L}
  -\ve^{-1}\sum_{\a=1}^3\ps{\p_\a (U^\ve_t G),\cl h_\a(\n_t)}_{0,L}\cr &
  -\ve^{-2}\sum_{\a=1}^3\ps{\nabla_\a(U^\ve_t G),
  \chi\nabla_\a \tilde I(\n_t(0))+w^{(a)}_{0,\a}(\n_t)}_{0,L}
+R_5(t,G,h),\cr }
\Eq(3.8)
$$
where, remembering that $\partial_tU^\ve_t=-\ve^{-1}E^*U^\ve_t$,
$$
R_5(t,G,h)=\sum_{\a=1}^3\ps{\p_\a
(-E^*U^\ve_t) G,h_\a(\n_t)}_{0,L}.
$$
Now, given $\bar D_{\a,\g}=(\bar D^{\b,\nu}_{\a,\g})_{\b,\nu=0,\ldots ,4}$,
$\a,\g=1,2, 3$, we add and subtract the term
$$
\sum_{\a,\g=1}^3\ps{ \bar D_{\a,\g}\p_\a(U^\ve_t
G),\nabla_\g \tilde I(\n_t(0))}_{0,L} 
%\Eq(3.10)
$$
in \equ(3.8). Then $\g^\ve_1(t,G,h)$ is equal to
$$
\eqalign{ & \ps{\p_t(U^\ve_t G),\tilde I(\n_t(0))}_{0,L}
  +\ps{ {\cal D}^*(U^\ve_t G), \tilde I(\n_t(0))}_{0,L}\cr &
  -\ve^{-1}\sum_{\a=1}^3\Bl\langle\p_\a (U^\ve_t G),
  w^{(a)}_{0,\a}(\n_t)-c_\a
  -\sum_{\g=1}^3\bar D_{\a,\g}\nabla_\g
  \tilde I(\n_t(0)) -\cl h_\a(\n_t)\Br\rangle_{0,L}\cr &
  +R_5(t,G,h)+R_1(t,G)+R_2(t,G)+R(t,G),\cr } \Eq(3.11)
$$
with ${\cal D}^*$ the adjoint in $L^2(\T_3,\R^5)$ of the differential
operator ${\cal D}$ in \equ(D), 
$c_\a=(c_\a^\b)_{\b=0,\ldots,4}$ the
equilibrium value of the current $w^{(a),\b}_\a$ (see \equ(3.7bis)) and
$$
R_1(t,G)=\ve^{-1}\sum_{\a=1}^3
\ps{(\p_\a -\ve^{-1}\nabla_\a) (U^\ve_t G), \chi\nabla_\a \tilde
  I(\n_t(0))}_{0,L},
$$
$$
R_2(t,G)=\sum_{\a,\g=1}^3\ps{
(\ve^{-1}\nabla^-_\g -\p_{\g}) \p_\a D_{\a,\g}(U^\ve_t
  G), \tilde I_\nu(\n_t(0))}_{0,L},
$$
$$
R(t,G)=\ve^{-1}\sum_{\a=1}^3\ps{\left(\p_\a
  -\ve^{-1}\nabla_\a\right) (U^\ve_t G),
w^{(a)}_{0,\a}-c_\a}_{0,L}.
$$
From the definition of $U^\ve_t G$, the first term of the sum
\equ(3.11) can be written as
$$\ve^{-1}\sum_{\a=1}^3
  \ps{(-E^* U^\ve_t) G,\tilde I(\n_t(0))}_{0,L}
  =\ve^{-1}\sum_{\a=1}^3
  \ps{\p_\a(U^\ve_t G),
  d_\a \tilde I(\n_t(0))}_{0,L} 
%\Eq(3.13)
$$
where the coefficients of the matrix
$d_\a=(d^{\b,\nu}_\a)_{\b,\nu=0\ldots,4}$ were defined in \equ(3.7d). 
Recalling the definition of the local functions 
$g_\a=(g^\b_{\a})_{\b=0\ldots,4}$ and 
$u_\a=(u^\b_{\a})_{\b=0\ldots,4}$ in \equ(3.13a), \equ(3.14)
(we omit the label $q$),  we obtain
$$
\g^\ve_1(t,G,h) =
  \ps{{\cal D}^*(U^\ve_t G),\tilde I_\nu(\n_t(0))}_{0,L}
   +\sum_{i=1}^4R_i(t,G)+\sum_{i=5}^6R_i(t,G,h),
$$
where 
$$
R_3(t,G)=
  \ve^{-1}\sum_{\a=1}^3\Bl\langle\Bl(\p_\a
    -\ve^{-1}\nabla_\a+{1\over 2}\nabla^-_{e_\a}\partial_\alpha\Br)
  (U^\ve_t G),d_\a\tilde I(\n_t(0))\Br\rangle_{0,L},
$$
$$
R_4(t,G)=\ve^{-1}\sum_{\a=1}^3\ps{(\p_\a
  -\ve^{-1}\nabla_\a) (U^\ve_t G), g_{\a}(\n_t)}_{0,L}
$$
and
$$
R_6(t,G,h)=-\ve^{-1}\sum_{\a=1}^3\ps{\p_\a
(U^\ve_t G),u_{\a}(\n_t)}_{0,L}.
$$
To prove that the compensator $\int_0^t\g^\ve_1(t,G,h)ds$ is converging,
%$$\limsup_{\ve\to
%  0}E^\ve\left[\left(\int_0^t\g^\ve_1(t,G,h)ds\right)^2\right]=0,\quad
%\b=0,\dots,4$$
we have to control the remainder terms. 

The remainders for $i=1,2,3$ are easily controlled by the following 
\vskip.2cm
\noindent{\bf \Lemma (s4444)} 
{\it Let $h$ be a mean zero local function and $G:\Bbb R_+\times \Bbb T^3\to \Bbb R$ a continuous
function. Then there exist a constant $c$ depending only on $h$ such
that, for all $t\ge 0$ and all
$\ve >0$
$$\Bbb E^\mu_\e\left[\sup_{0\le t\le T}\left(\int_0^t\!ds\,
 \ve^{3/2}\sum_{x} G(s, \e x)\tau_x
h(\eta_s)\right)^2\right]\le c\,T^2\|G\|^2_\infty.$$ }
\vskip.2cm
The proof is an easy consequence of the Schwartz inequality  
and the stationarity of $\P_\ve^\mu$. We refer to Lemma 4.1 in [CLO] for details.

By Taylor expanding and using Lemma \equ(s4444) we immediately obtain
$$
\lim_{\ve\to 0}\E^\mu_\ve\left[\left(
      \int_0^t\!ds\,R_i(s,G)\right)^2\right]= 0.
$$
for $i=1,2,3$.

The other terms are estimated by using the refined Boltzmann-Gibbs
principle (Theorem \equ(s2.3)) because the functions $g^\b_\a$
and $u^\b_\a$ are in $\cg$ ($h^\b_\a\in \cg$ by hypothesis). We get
$$
\lim_{\ve\to 0}\E^\mu_\ve\left[\left( \int_0^t\!ds\,R_4(s,G)\right)^2\right]
=\lim_{\ve\to 0}\E^\mu_\ve\left[\left(
    \int_0^t\!ds\,R_5(s,G,h)\right)^2\right]= 0 
%\Eq(3.20)
$$
and
$$
\limsup_{\ve\to 0}\E^\mu_\ve\left[\left(
    \int_0^t\!ds\,R_6(s,G,h)\right)^2\right]\le
c\,t\max_{\a=1,2,3}
\|\partial_\a(U^\ve_\cdot G)\|^2_0
\sum_{\a=1}^3\sum_{\b=0}^4V(u^\b_\a;r,\theta). 
%\Eq(3.20a)
$$
From the definition of the semi-group $(U^\ve_t)$, it is clear that
$\|\partial_\a(U^\ve_\cdot G)\|^2_0$ is uniformly bounded in $\ve$. 
Moreover the diffusion coefficients $\bar D^{\b,\nu}_{\a,\g}$ are chosen
in such a way that, since we take for $h^\b_\a$ the terms of the sequence
$(h^{(q),\b}_\a)$ given in Theorem \equ(so3.1b), we have
$$
\lim_{q\to \infty}V(u^{(q),\b}_\a;r,\theta)=0.
%\Eq(3.21a)
$$
We have shown so far that there exists a random variable $R^{q}_\ve$
which converges to $0$ in $L^2(\P^\mu_\ve)$ as $\ve\to 0$ and then $q\to
\infty$ such that
$$
M^\ve_1(t,G,h)
=\xi^\ve\bl(t,G\br)-\xi^\ve\bl(0,G\br)-\int_0^t\! ds\,\ps{{\cal D}^*(U^\ve_s
G),\tilde I(\n_s(0))}_{0,L}+R^{q}_\ve.\Eq(?)
$$
We would like to
have instead of the third term in \equ(?)  a term of the form 
$$
\ps{U^\ve_s(HG),\tilde I(\n_s(0))}_{0,L}
$$
for some suitable operator $H$ that we
could then rewrite  as $\xi^\ve(s,H G)$, so to identify the limiting martingale
problem. We proceed in the following way:
$$
\ps{{\cal D}^*(U^\ve_sG),\tilde I(\n_s(0))}_{0,L}= 
\ps{U^\ve_s (U^\ve_s) ^{-1} {\cal
  D^*}U^\ve_s G,\tilde I(\n_s(0))}_{0,L}.
$$
% By Lemma A.1 and Lemma A.2
% $$
% \lim_{\e\to 0}||(U^\ve)_s ^{-1} {\cal D}U^\ve_s G||_2=||\Pi_E({\cal
%   D})G||_2$$
By Lemma A.2 and Lemma A.3
$$
\lim_{\ve\to 0}\E^\mu_\ve\lt[\lt(\int_0^t\! ds\,
 \Bl[\xi^\ve\bl(s,(U^\ve_s)^{-1}
{\cal D}^*U^\ve_s G\br)-\xi^\ve\bl(s,\pi_{E^*}({\cal D}^*) G\br)\Br]
\rt)^2\rt]=0.
$$
Hence, noticing that $\pi_{E^*}({\cal D}^*)=\pi_{E}({\cal D})^*$, we
have proved that there exists a random
variable $C_\ve^q$ which converges to $0$ in $L^2(\P^\mu_\ve)$ when $\ve\to
0$ and then $q\to\infty$ such that
$$
M^\ve_1(t,G,h)
=\xi^\ve(t,G)-\xi^\ve(0,G)-\int_0^t\!ds\,\xi^\ve\bl(s,
  \pi_E({\cal D})^* G\br)+C_\ve^q.
$$
\vskip .5truecm

We now compute the compensator $\g^\ve_2$ in \equ(3.2). We first
remark that 
$$
\g^\ve_2(t,G,h) =
\bigl(\ve^{-2}\cl\bigr)\bigl(\z^\ve(t,G,h)^2\bigr)
-2\z^\ve(t,G,h)
\bigl(\ve^{-2}\cl\bigr)\z^\ve(t,G,h).
%\Eq(3.21)
$$
We introduce the operator $\cl^{(2)}=\cl^{{\rm ex},(2)}+\cl^{{\rm c},(2)}$ for
local functions $f$ and $g$ as
$$
\eqalign{
\cl^{{\rm ex},(2)}(f;g) &=\cl^{\rm ex}(fg)-f\cl^{\rm ex} g-g\cl^{\rm ex} f, \cr
\cl^{{\rm c},(2)}(f;g) &=\cl^{\rm c}(fg)-f\cl^{\rm c} g-g\cl^{\rm c} f. \cr
}
%\Eq(3.22)
$$
Then we obtain
$$
\g^\ve_2(t,G,h) = Y^\ve_1(t,G)+Y^\ve_2(t,G,h)+
Y^\ve_3(t,G,h), 
%\Eq(3.23)
$$
where
$$
\eqalign{ Y^\ve_1(t,G) &= \ve\sum_{\b,\nu=0}^4\sum_{x,y}(U^\ve_t
  G)_{\b}(\ve x)(U^\ve_t G)_{\nu}(\ve y)
  \cl^{{\rm ex},(2)}\bigl(\tilde I_\b(\n_x);\tilde I_\nu(\n_y)\bigr),\cr
  Y^\ve_2(t,G,h)&=
  -2\ve^2\sum_{\a=1}^3\sum_{\b,\nu=0}^4\sum_{x,y}(U^\ve_t G)_{\b}
  (\ve x)\p_\a (U^\ve_t G)_{\nu}(\ve y)\cl^{{\rm ex},(2)}
  \bigl(\tilde I_\b(\n_x);\tau_yh^\nu_\a)\bigr),\cr
  Y^\ve_3(t,G,h)&=\ve^3\sum_{\a,\g=1}^3\sum_{\b,\nu=0}^4\sum_{x,y}
  \p_\a (U^\ve_t G)_{\b}(\ve x)\p_\g (U^\ve_t G)_{\nu}(\ve
  y)\cl^{(2)}
  \bigl(\tau_xh^\b_\a;\tau_yh^\nu_\g)\bigr).\cr 
} 
%\Eq(3.24)
$$
From the explicit formulas \equ(3.25), \equ(3.27) and \equ(3.28) that
we will get below for $Y_i(t,G)$, $i=1,2,3$ and the use of Lemma \equ(s4444) it is
easy to see that
$$
\int_0^t\!ds\,\Bl(\g^\ve_2(s,h,G)-
\E^\mu_\ve\bl[\g^\ve_2(s,h,G)\br]\Br)
$$
is converging to 0 in $L^2(\P^\mu_\ve)$. So,  all we need to compute  is
$\E^\mu_\ve\bl[\g^\ve_2(t,h,G)\br]$.

\bigskip

Notice that
$$
\eqalign{
\cl^{{\rm ex},(2)}(f;g)& =\sum_{x,e,v}b(x,x+e,v)
\nabla_{x,x+e,v}f\nabla_{x,x+e,v}g, \cr
\cl^{{\rm c},(2)}(f;g)& =\sum_{x,q}
\nabla_{x,q}f\nabla_{x,q}g, \cr
}
\Eq(3.24a)
$$
with $\nabla_{x,x+e,v}f=f(\n^{x,x+e,v})-f(\n)$,
$\nabla_{x,q}f=f(\n^{x,q})-f(\n)$ and
$$
b(x,y,v)=\Bl(\chi+{1\over 2}v\cdot (y-x)\Br)
\eta(x,v)(1-\eta(y,v)).
$$
So, if we let $\phi_0(v)=1$, $\phi_\b(v)=v_\b$ for $\b=1,2,3$ and
$\phi_4(v)={1\over 2}|v|^2$, a straightforward computation leads to
the following
$$
\eqalign{ Y^\ve_1(t,G) = &
  \ve^3\sum_{\a=1}^3\sum_{\b,\nu=0}^4\sum_{x}\Bl(\pt_\a (U^\ve_t
  G)_{\b}\pt_\a (U^\ve_t G)_{\nu}\Br)(\ve x)\times\cr
  &\qquad\sum_v\Bl[b(x,x+e_\a,v)+b(x+e_\a,x,v)\Br]\phi_\b(v)\phi_\nu(v)+O(\ve).}
\Eq(3.25)
$$
Therefore
$$
\eqalign{ \E^\mu\Bigl[Y^\ve_1(t,G)\Bigr]&=\ve^3
  2\chi\sum_{\a=1}^3\sum_{\b,\nu=0}^4\sum_x C_{\b,\nu}\Bigl(\p_\a (U^\ve_t
  G)_{\b}\p_\a (U^\ve_t G)_{\nu}\Bigr)(\ve x)
  +O(\e)\cr &=-2\chi\ps{(U^\ve_t G),\Delta C(U^\ve_t
  G)}_0+O(\e),} 
%\Eq(3.26)
$$
where $\Delta$ is the vectorial Laplacian operator defined as
$(\Delta G)_\b=\Delta G_\b$. Observe that
$$
\ps{(U^\ve_t G),\Delta C(U^\ve_t G)}_0
=\ps{G,e^{-{t\over \ve}E}Ce^{-{t\over \ve}E^*} \Delta G}_0
=\ps{G,Ce^{{t\over \ve}E^*}e^{-{t\over \ve}E^*}\Delta G}_0,
$$
where we have used that
$EC=-CE^*$. In conclusion, $Y^\ve_1(t,G)$ converges in $L^2(\P^\mu_\ve)$ to
$-2\chi \ps{G,C\Delta G}_0$.
% $$
% 2\chi\sum_{\a=1}^3\int_{\T_3}\!dx\, 
% \pt_\a G(x)\cdot C\pt_\a G(x)
% $$

\bigskip

We get in the same way
$$
\eqalign{ Y^\ve_2(t,G,h) = & 2\ve^3\sum_{\a,\g=1}^3\sum_{\b,\nu=0}^4
  \sum_{x}\Bl(\partial_\a (U^\ve_t G)_{\b}
  \partial_\g (U^\ve_t G)_{\nu}\Br)(\ve x)
  \times\cr
  &\qquad\sum_v\Bigl[b(x,x+e_\a,v)-b(x+e_\a,x,v)\Bigr]
  \phi_\b(v)\nabla_{x,x+e_\a,v}\Gamma(h_\a^\nu)+O(\ve),
  } \Eq(3.27)
$$
where $\Gamma(h_\a^\nu)=\sum_x \tau_xh_\a^\nu$. Since $\mu$ is
invariant for the jump
generator of particles with a given velocity, it is easy to check that
$$
\E^\mu\Bl[\n(x,v)(1-\n(x+e_\a,v))\nabla_{x,x+e_\a,v}\Gamma(h_\a^\nu)\Br]
=0,
$$
which implies that the time integral of $Y^\ve_2(t,G,h)$ converges to $0$ in $L^2(\P^\mu_\ve)$
by Lemma \equ(s4444).

\bigskip

The last term $Y^\ve_3(t,G,h)$ is given by
$$
\eqalign{
& \ve^3\sum_{\a,\g=1}^3\sum_{\b,\nu=0}^4\sum_{x}\Bl(\partial_\a (U^\ve_t
  G)_{\d} \partial_\g (U^\ve_t G)_{\nu}\Br)(\ve
  x)\times\cr 
&\qquad \left[ \sum_{v;|e|=1}b(x,x+e,v)
  \nabla_{x,x+e,v}\Gamma(h_\a^\b)\nabla_{x,x+e,v}\Gamma(h_\g^\nu)
  +\sum_{q}\nabla_{x,q}\Gamma(h_\a^\b)\nabla_{x,q}\Gamma(h_\g^\nu)\right].
  } \Eq(3.28)
$$
By using again Lemma \equ(s4444) it is immediate to show that the time integral of $Y^\ve_3$
converges in $L^2(\P_\ve^\mu)$ to its average that we are going to compute.

Let $\cl_s^{\rm ex}$ be respectively the symmetric part of $\cl^{\rm ex}$ 
in
$L^2(\mu)$. It is easy to check that for any local function $f$ and $g$
$$
  \sum_{v;|e|=1}\E^\mu\Bigl[\eta(0,v)(1-\eta(e,v))\nabla_{0,e,v}\Gamma(f)
  \nabla_{0,e,v}\Gamma(g)\Br]=
  2\E^\mu\Bigl[\G (f)(-\cl_s^{\rm ex})g\Br]
$$
and
$$
\sum_{q}\E^\mu\Bigl[\nabla_{0,q}\Gamma(f)
  \nabla_{0,q}\Gamma(g)\Br]=
  2\E^\mu\Bigl[\G (f)(-\cl^{\rm c})g\Br].
$$
Therefore
$$
\E^\mu\bl[Y^\ve_3(t,G)\br]=2\ve^3\sum_x\E^\mu\Bl[\G\bl(\partial(U^\ve_t G)(\ve
x)\cdot h\br)(-\cl_s)(\partial(U^\ve_t G)(\ve x)\cdot h)\Br],
$$
where $\cl_s$ and $a\cdot b$ were defined after \equ(4.15). Remember that the
functions $h=(h_\a^\b)$ are chosen as the terms of the sequence
$(h_\a^{(q),\b})$ in Theorem \equ(so3.1b).
Lemma \equ(so4.6) asserts that
$$
\lim_{q\to\infty}\E^\mu\Bl[\G\bl(a\cdot h^{(q)}\br)(-\cl_s)(
a\cdot h^{(q)})\Br]
=2a\cdot(\bar DC)a.
$$
Hence,
$$
\eqalign{ 
\E^\mu\bl[Y^\ve_3(t,G,h)\br]&=
2\ve^3\sum_x\underline{\partial}(U^\ve_t G)\cdot (\bar DC)
\underline{\partial}(U^\ve_t
G)+o_q(1)\cr 
&=-2\ps{U^\ve_t G,(\bar{\cal D}C)(U^\ve_t G)}_0 +o_q(1)+O(\ve),
}
$$
where, denoting by $\bar D_{\a,\g}$ the matrix $(\bar D_{\a,\g}^{\b,\nu})_{\b,\nu=0\ldots,4}$,
$$
\bar{\cal D}G=\sum_{\a,\g=1}^3
\bar D_{\a,\g}\pt_\a\pt_\g G.
$$

With the property $\bar{\cal D}C=C\bar{\cal D}^*$, we get
$$
\ps{U^\ve_t G,(\bar{\cal D}C)(U^\ve_t G)}_0
=\ps{e^{-{t\over \ve}E^*} G,(C\bar{\cal D}^*)e^{-{t\over \ve}E^*}G}_0
=\ps{ G,Ce^{-{t\over \ve}E^*}\bar{\cal D}^*e^{-{t\over \ve}E^*}G}_0
$$
and by Lemma A.2
$$
\lim_{\ve\to 0}\int_0^t\!ds\,\exp\bl({s\over\ve }E^*\br) \bar{\cal D}^*\exp
\bl(-{s\over\ve }E^*\br)=t\,\pi_{E^*}(\bar{\cal D}^*)=t\, \pi_{E}(\bar{\cal D})^*,
$$
so
$$
\lim_{\ve\to 0}\E^\mu\Bl[\int_0^t\!ds\,Y^\ve_3(s,G,h)\Br]
=-2t\,\ps{G, C\,\pi_{E}(\bar{\cal D})^* G}_0.
$$
To summarize, we have proved that there exists a random variable
$R^q_\ve$ vanishing in $L^2(\P^\mu_\ve)$ in the limits $\ve\to 0$ and
then $q\to\infty$ such that
$$
\eqalign{
M_2^\ve(t,G,h)&=\bl(M_1^\ve(t,G,h)\br)^2 +2t\ps{G, C\,
\pi_{E}(\bar{\cal D})^* G}_0+2t\ps{G, C\,\Delta G}_0+R^q_\ve\cr
&=\bl(M_1^\ve(t,G,h)\br)^2 +2t\ps{G, C\,\pi_{E}({\cal D})^*G}_0+R^q_\ve.
}
$$
This completes the proof of Theorem \equ(so2.2), once the Boltzmann-Gibbs 
principle and Lemmas A.2
and A.3 are proved.

\vskip 1truecm
\noindent{\bf 5. The Boltzmann-Gibbs principle.}
\vskip .5truecm \numsec= 5 \numfor= 1 \numtheo=1

Since we closely follow the strategy proposed in [CLO] 
to prove Theorem \equ(s2.3),
we will only focus our attention to the points where
non trivial changes are necessary.

One of the ingredients in the proof is the equivalence of
ensembles, which is classical for Bernoulli product measures but, as
far as we know, is not in our case. 
We state below a weaker
statement which will suffice to our purpose.

For a given chemical potential $n\in\R^5$, let
$M(n)=(M_0(n),\ldots,M_4(n))$ be defined as 
$M_\b(n)=\E^{\mu_{L,n}}[I_\b(\n_0)]$. If we put $A=M(\R^5)$, it is easy
to verify that $n\mapsto M(n)$ is a $C^1$ diffeomorphism from $\R^5$
onto $A$, in particular the inverse function $M\mapsto n(M)$ is
continuous on $A$. Given
$a>0$, we introduce the set $A^a$ of $M\in A$ such that,
$|n(M)-n_0|\le a$, with $n_0=(r,0,0,0,\t)$ the
equilibrium chemical potentials.
We denote by $\bar\mu_{L,M}$ the grand canonical
measure $\mu_{L,n(M)}$ which satisfies therefore 
$E^{\bar \mu_{L,M}}[I_\b(\n_0)]=M_\b$ for $\b=0,\ldots,4$.

Recall that $\bar I_L^+(\n)=\bl(\bar I_{0,L}^+(\n),\ldots, 
\bar I_{4,L}^+(\n)\br)$ are the empirical averages of the conserved
quantities in $\L_L$.
For any particle configuration $\n$ in $\O_L$, we call $\bar N^v_L(\n)$,
$v\in\cv$, the average number of particles with velocity $v$ in $\L_L$.

Also recall the definition of $\phi_\b(v)$ before \equ(3.24a). 
Given $k=(k_v)_{v\in\cv}$, we set
$I_\b(k)=\sum_v\phi_\b(v)k_v$ and $I^+(k)=\bl(I_0(k),\ldots,I_4(k)\br)$.

\medskip
\goodbreak

\noindent{\bf \Lemma (lbg0) (Equivalence of ensembles)}
{\it
Let $h$ be a local function.
Then there exists a constant $c=c(h,a)$ such that
$$
\Bigl|\E^\mu\bigl[h\big|\bar I^+_L=M\bigr]-E^{\bar\mu_{L,M}}[h]\Bigr|\le c\,\ve^{3}
$$
\nobreak
uniformly in $M\in A^a$.
}

\smallskip
\noindent{\it Proof.} Let $\ell$ be the number of velocities in $\cv$ 
and denote by $\nu_{\a}$,
$\a=(\a_v)_{v\in\cv}$, the product measure on $\O_L$ of Bernoulli
measures with parameters $\a=(\a_v)_v$, i.e. $E^{\nu_{\a}}[\n(x,v)]=\a_v$. A
straightforward extension of the classical strong equivalence of ensembles
asserts that for any local function $g$, 
$$
\Bigl|E^{\nu_\a}\bigl[g\big|\bar N^v_L=k_v,v\in\cv\bigr]-E^{\nu_k}[h]\Bigr|\le
C(h)\ve^3
\Eq(clt)
$$
uniformly in $k=(k_v)_{v\in\cv}\in B_L=\{0,L^{-3},\ldots,1\}^\ell$.

We first compute the term $\E^\mu\bigl[h\big|\bar I^+_L=M\bigr]$. 
Since this expectation 
does not depend on the chemical potential (here $n_0$),
it is equal to $E^{\nu_{1/2}}\bigl[h\big|\bar I^+_L=M\bigr]$
with the obvious abuse of notation $1/2=(1/2,\ldots,1/2)$.
Therefore, from \equ(clt),
$$
\E^\mu\bigl[h\big|\bar I^+_L=M\bigr]=
{
\sum_{k\in B_L, \bar I^+_L(k)=M}\nu_{1/2}\bl(\bar N^v_L=k_v,v\in\cv\br)
E^{\nu_k}[h]
\over
\sum_{k\in B_L,\bar I^+_L(k)=M}\nu_{1/2}\bl(\bar N^v_L=k_v,v\in\cv\br)
}+O(\ve^3).
\Eq(f1)
$$
Since the particles with different velocities are independent
$$
\nu_{1/2}\bl(\bar N^v_L=k_v,v\in\cv\br)=\prod_{v\in\cv}
\nu_{1/2}\bl(\bar N^v_L=k_v\br)
$$
and the asymptotics of a single term in the product above
is given by the Stirling formula
$$
\nu_{1/2}\bl(\bar N^v_L=k\br)=
{1\over \sqrt{2\pi \ve^{-3}k(1-k)}}\exp\Bl[-\ve^{-3}(s(k)+\log 2)\Br]
\Bl(1+O\Bl({\ve^3\over k(1-k)}\Br)\Br),
\Eq(f2)
$$
where $s(k)=k\log k+(1-k)\log(1-k)$ is the entropy. 
In particular, if $(k_v)_v$ belongs to $B_L^\d:=B_L\cap[\d,1-\d]^\ell$ 
for some small $\d>0$, then
$$
\nu_{1/2}\bl(\bar N^v_L=k_v,v\in\cv\br)=
{1\over \sqrt{(2\pi \ve^{-3})^\ell\prod_vk_v(1-k_v)}}
\exp\bl[-\ve^{-3}\sum_v(s(k_v)+\log 2)\br]
\bl(1+O(\ve^3)\br).
$$
The fact that the entropy is  convex suggests to use the Laplace method to
derive the asymptotics of both terms in the ratio \equ(f1). This is
the aim of Lemma \equ(lbg0a) below which is stated in the $\ell=1$
case without any constraint on $k$,
nevertheless the generalization to higher dimension with constrains is easy
because, up to a linear change of variables $k\mapsto k'$, the sums
over $k$ in \equ(f1) with constraints
can be written as a sum without constraint 
over $k'$ in a cube of dimension $\ell-5$
(5 is the number of linear conditions $I^+(k)=M$). 
Therefore, we have 
$$
\eqalign{ &\sum_{k\in B^\d_L,
  \bar I^+(k)=M}\nu_{1/2}\bl(\bar N^v_L=k_v,v\in\cv\br)E^{\nu_k}[h]\cr
  &\qquad={T_\ve\over \sqrt{(2\pi \ve^{-3})^\ell\prod_vk^*_v(1-k^*_v)}}
  \exp\Bl[-\ve^{-3}\sum_v(s(k^*_v)+\log
  2)\Br]E^{\nu_{k^*}}[h]\bl(1+O(\ve^3)\br),\cr
}
$$
where $k^*$ is the minimizer of $\sum_v(s(k_v)+\log 2)$ under the
constraints $k\in [\d,1-\d]^n$ and $I^+(k)=M$,
$$
T_\ve=\sum_{\scriptstyle k\in B^\d_L,\bar I^+(k)=M\atop\scriptstyle
|k-k^*|\le \ve^{3\a}}\exp\Bl[-{\sum_vs''(k^*_v)\over
2}\ve^{-3}(k-k^*)^2\Br].
$$
with $0<\a<1/2$. Notice that this result holds 
provided that the minimizer $k^*$ satisfies
$k^*\in ]\d,1-\d[^\ell$, that will be shown below. 
As a consequence, the ratio \equ(f1) is equal to
$$
E^{\nu_{k^*}}[h]\bl(1+O(\ve^3)\br),
$$
provided that the contributions from ``bad'' configurations are
negligible.

Let $\k$ be the minimizer of $\sum_v(s(k_v)+\log 2)$ under the
constraints $k\in [0,1]^\ell$, $I^+(k)=M$.
From Lagrange optimization theorem,
$\k$ has to minimize the function
$$
\sum_v(s(k_v)+\log 2)+\sum_{\b=0}^4\g_\b
\sum_v\br(\phi_\b(v)k_v-M_\b\bl),
$$
where $\phi_\b$ have been defined in the line before \equ(3.25) 
and $(\g_\b)$ are Lagrangian multipliers.
So the minimizer satisfies
$$
s'(\k_v)=\sum_{\b=0}^4\g_\b\phi_\b(v),\quad v\in\cv.
$$
Since the derivative of the entropy $s'(\a)$ is equal to the
associated chemical potential $\l=\log{\a\over 1-\a}$, we have
$\nu_{\k}=\mu_{L,\g}$, $\g=(\g_0,\ldots,\g_4)$ but the
constraint $I^+(\k)=M$ implies that $\g=n(M)$ that is to say 
$\nu_{\k}=\bar\mu_{L,M}$ and in particular
$E^{\nu_{\k}}[h]=E^{\bar\mu_{L,M}}[h]$.
Moreover, if $\l_v=\log{\k_v\over 1-\k_v}$ is the chemical 
potential related to $\k_v$, then we have 
$$
\l_v=\sum_{\b=0}^4\phi_\b(v)n_\b(M).
$$
From the assumption $M\in A^a$, the previous equality implies that we
can choose $\d>0$ small enough such that $\k\in]2\d,1-2\d[^\ell$
uniformly in $M\in A^a$. Such a choice of $\d$ implies that
$k^*=\k$.

So the lemma will be proved if we finally show that the contribution of the
``bad'' $k$ ($k\in B_L\setminus [\d,1-\d]^\ell$) inside the sums in the
numerator and denominator of the ration \equ(f1) is irrelevant with
respect to the leading term. From Stirling formula \equ(f2), there is
$c>0$ such that
$$
\sum_{k\notin B^\d_L,
  \bar I^+(k)=M}\nu_{1/2}\bl(\bar N^v_L=k_v,v\in\cv\br)
\le \exp\bl[-\ve^{-3}\sum_v(s(k_v)+\log 2)-c\log\ve\br].
$$
From the discussion above, there exists $b>0$ such that 
$\sum_vs(k_v)\ge \sum_vs(\k_v)+b$, therefore
$$
\sum_{k\notin B^\d_L,
  \bar I^+(k)=M}\nu_{1/2}\bl(\bar N^v_L=k_v,v\in\cv\br)
\le c\exp\bl[-\ve^{-3}\sum_v(s(\k_v)+\log 2)\br]\exp[-b\ve^{-3}/2].
$$
\qed

\bigskip

\noindent{\bf \Lemma (lbg0a)} 
{\it
Let $\psi$ and $\phi$ be smooth functions
on $[0,1]$, $\psi$ concave, $\phi$ non negative. Assume that
the maximizer $\t$ of $\psi$ is in $]0,1[$, then
$$
\sum_{i=0}^N\phi\bl({i\over N}\br)\exp \bl[N\psi\bl({i\over N}\br)\br]
=S_N\bl(\a,{\psi''(\t)\over 2}\br)\phi(\t)\exp [N\psi(\t)]
\Bigl(1+O\bl({1\over N}\br)\Bigr),
$$
with $\t$ the maximizer of $\psi$ and
$$
S_N(\a,a)=\sum_{|i-N\t|\le N^{1-\a}}\exp\Bl[a{(i-N\t)^2\over N}\Br], 
\ 0<\a<{1\over 2}.
%\ ,\qquad\bl|O_\phi\bl({1\over N}\br)\br|\le {c(\phi)\over N}
$$
}

\medskip
\noindent{\it Proof.} We start by factorizing the leading term 
$\exp[N\psi(\t)]$ in the sum. For simplicity call
$$
U_N(i)=\phi\bl({i\over N}\br)\exp N\bl[\psi\bl({i\over
N}\br)-\psi(\t)\br].
%\Eq(le1)
$$
From the assumption on $\t$, if $\d>0$ there exists a constant $c(\d)>0$ such that
$$
\sum_{|i-N\t|> N\d}U_N(i)\le \exp[-Nc(\d)].
$$
Moreover, choosing $\d$ small enough ensures that $\psi(x)-\psi(\t)\le
-c(x-\t)^2$ provided that $|x-\t|\le\d$, where $c>0$ is a constant
which will change from line to line. Then, given $0<\a<{1\over 2}$,
$$
\sum_{N^{1-\a}<|i-N\t|\le N\d}U_N(i)\le \exp[-N^{1-2\a}c(\d)].
\Eq(le2)
$$
So the main contribution is coming from $\sum_{|i-N\t|\le
  N^{1-\a}}U_N(i)$.
Using Taylor expansion, we see that in this range of $i$'s,
$$
U_N(i)=\exp\Bl[b_0{j^2\over N}\Br]\lt(
a_0+a_1\Bl({j\over N}\Br)+a_0b_1N\Bl({j\over N}\Br)^3
+O\Bl(\Bl({j\over N}\Br)^2\Br)+O\Bl(N\Bl({j\over N}\Br)^4\Br)\rt),
$$
where $j={i\over N}-\t$, $a_0=\phi(\t)$, $a_1=\phi'(\t)$, 
$b_0={\psi''(\t)\over 2}<0$ and
$b_1={\psi^{(3)}(\t)\over 6}$. By the ``almost oddness'' of $j$,
$$
\sum_{|i-N\t|\le N^{1-\a}}\exp\Bl[b_0{j^2\over N}\Br]\lt(
a_1\Bl({j\over N}\Br)+a_0b_1N\Bl({j\over N}\Br)^3\rt)
=O\bl({1\over N}\br)S_N(\a,b_0).
\Eq(le3)
$$
We also remark
$$
\sum_{|i-N\t|\le N^{1-\a}}\Bl({j\over N}\Br)^2\exp\Bl[b_0{j^2\over
  N}\Br]
\le {c\over N}\sum_{|i-N\t|\le N^{1-\a}}\exp\Bl[b_0{j^2\over
  2N}\Br],
$$
so that 
$$
\sum_{|i-N\t|\le N^{1-\a}}\Bl({j\over N}\Br)^2\exp\Bl[b_0{j^2\over
  N}\Br]=O\bl({1\over N}\br)S_N\bl(\a,{b_0\over 2}\br),
\Eq(le4)
$$
We get in the same way
$$
\sum_{|i-N\t|\le N^{1-\a}}N\Bl({j\over N}\Br)^4\exp\Bl[b_0{j^2\over
  N}\Br]=O\bl({1\over N}\br)S_N\bl(\a,{b_0\over 2}\br).
\Eq(le5)
$$
Finally, comparing $N^{-1/2}S_N(\a,a)$ ($a<0$) with the integral of a
Gaussian, it is easy to check that $S_N(\a,a)=c(a)\sqrt{N}+O(1)$,
therefore 
$$
S_N\bl(\a,{b_0\over 2}\br)=O(1)S_N(\a,b_0).
\Eq(le6)
$$
Putting together formulas \equ(le2) to \equ(le6), the Lemma is proved.

\noindent\qed

\bigskip

Even if the equivalence of ensembles that we stated in Lemma \equ(lbg0) is
weaker than the classical one, it is enough to prove the following
result which is actually the only estimate needed in the proof of  
 Boltzmann-Gibbs principle.

\bigskip

\noindent{\bf \Corollary (lbg0b)} 
{\it If $h\in\cg$ is a local function then 
$$
\E^\mu\Bl[\bl(\E^\mu\bigl[h\big|\bar I^+_L=M\bigr]\br)^2\Br]\le c\,\ve^6.
$$
}

\smallskip
\noindent{\it Proof.} Let $\hat h=\E^\mu\bigl[h\big|\bar I^+_L\bigr]$,
$\tilde h(M)=E^{\bar\mu_{L,M}}[h]$ and consider the decomposition 
$$
\E^\mu\bl[\hat h^2\br]\le
2\E^\mu\Bl[\bl(\hat h-\tilde h(\bar I^+_L)\br)^2\Br]+
2\E^\mu\Bl[\bl(\tilde h(\bar I^+_L)\br)^2\Br].
$$
Since $h$ is in $\cg$, we have 
$$\tilde h(m)=0  \quad\hbox{and } \quad{\pt \tilde h\over \pt
M_\b}\Big|_{M=m}=0,$$
with $m_\b$ the equilibrium values of $I_\b(\n_0)$. Therefore
$$
\bl|\tilde h(\bar I_L^+)\br|\le 
c\sum_{\b,\nu=0}^4\bl|\bl(\bar I_{\b,L}(\n)-m_\b\br)
\bl(\bar I_{\nu,L}(\n)-m_\nu\br)\br|.
$$
Hence
$\E^\mu_\ve\bl[(\tilde h(\bar I_L^+))^2\br]\le c\, \ve^6$.

On the other hand, it results from Lemma \equ(lbg0) that for any $a>0$
$$
\E^\mu\Bl[\bl(\hat h-\tilde h(\bar I^+_L)\br)^2\Br]
\le c(a)\ve^6+cP^\mu\bl[I^+_L\notin A^a\br].
$$
From the continuity of the function $M\mapsto n(M)$, there exists $b>0$
such that 
$$
P^\mu\bl[I^+_L\notin A^a\br]\le P^\mu\bl[|I^+_L-m|>b\br].
$$
Finally, since $I^+_L={1\over |\L_L|}\sum_xI^+(\n_x)$ with $I^+(\n_x)$
i.i.d. random vectors with finite exponential moments and expectation
$m$ under $\mu$, a large deviation estimate provides
$$
P^\mu\bl[|I^+_L-m|>b\br]\le\exp(-c\ve^3).
$$
\qed
\bigskip

The first result used in [CLO] (Lemma 4.3) before establishing
 Boltzmann-Gibbs is a general estimate
bounding the equilibrium expectation of the squared time integral of
zero mean functions of Markov processes by their $\|\cdot\|_{-1}$
norm.  More precisely, if $X$ is a Markov process on the finite state space
$\ce$ with generator $L$ and ergodic invariant measure $\pi$, then
there exists a (universal) constant $c>0$ such that for any
function $f\,:\,[0,T]\times\ce\rightarrow\R$ satisfying
$\E^\pi[f(t,X_t)]=0$ for any $t\in[0,T]$, we have

$$
\E^\pi\left[\sup_{0\le t\le T}\left(
\int_0^t\!ds\,f(s,X_s)\right)^2\right]
\le c\int_0^T\!ds\,\|f(s,\cdot)\|_{-1}^2,
\Eq(bg1)
$$
where
$$
\|f\|_{-1}^2=\sup_g\{\ps{f,g}+\ps{f,L^sg}\}
$$
and  $L^s$ is the symmetric part of $L$ in $L^2(\pi)$.

\bigskip

The next lemma (Lemma 4.4 in [CLO]) is needed to control remainder
terms in the proof of the Boltzmann-Gibbs principle.

\medskip
\noindent{\bf \Lemma (lbg1)}
{\it
For any local function $h\in\cg$, there exists a constant $c(h)>0$
such that for any subset $B$ of $\L_L$, any smooth function 
$G\,:\,[0,T]\times\T_3\rightarrow\R$ and $\ve$ small enough,
$$
\E^\mu_\ve\left[\sup_{0\le t\le T}\left(
\int_0^t\!ds\,\ve^{3/2-1}\sum_{x\in B}G(s,\ve x))
\tau_xh(\n_s)\,ds\right)^2\right]
\le c(h)(1+T)\int_0^T\!\!ds\,\ve^3\sum_{x\in B}G^2(s,\ve x).
\Eq(apriori)
$$
}

\noindent{\it Proof.} 
Following [EMY2] (section 4), we introduce an alternative
representation for the particle configuration
$\n_x=\bigl(\n(x,v)\bigr)_{v\in\cv}$ at  site $x$: one can find  $n(n-5)$
numbers $c_\b(v)$, $\b=-n+5,\ldots,-1$ and $v\in\cv$, such that if we
put
$$
I_\b(\n_x)=\sum_{v\in\cv}c_\b(v)\n(x,v),
%\Eq(bg2)
$$
then the map $\n_x\mapsto (I_\b(\n_x))_{\b=-n+5,\ldots,4}$ is one to one.
Moreover the coefficients $c_\b(v)$ can be chosen in such a way that
the covariances $\E^\mu[I_\b(\n_x);I_\nu(\n_x)]$, $\b\not=\nu$, vanish
(except $\b,\nu\in \{0,4\}$). We also introduce the variables
$\tilde I_\b(\n_x)=I_\b(\n_x)-\E^\mu[I_\b(\n_x)]$.

\medskip

Let $\cg^{ex}$ be the space of functions $h$ such that
$\E^\mu[h]=0$ and $\sum_x\E^\mu[h;I_\b(\n_x)]=0$ for any $\b=-n+5,\ldots,4$.
The integration by parts lemma valid for ASEP (Lemma 6.1 in [EMY1]) easily
generalizes to a superposition of ASEP.

%  and, in our setting, it has the
% following form: for $h\in\cg^{ex}$, there exist functions
% $\Phi_{x,e,v}\in L^2(\mu)$ (depending on $h$) such that for 
% any $u\in L^2(\mu)$,
% $$
% \E^\mu[gu]=\sum_x\sum_{|e|=1}\sum_{v\in\cv}\E^\mu[\Phi_{x,e,v}\nabla_{x,e,v}u]
% \Eq(bg3)
% $$
% with $\nabla_{x,e,v}u=u(\n^{x,x+e,v})-u(\n)$. Furthermore
% $$
% \sum_x\sum_{|e|=1}\sum_{v\in\cv}|x|^{3+1/2}\E^\mu[\Phi^2_{x,e,v}]<+\infty
% \Eq(bg4)
% $$

\smallskip
We now turn to the proof. Fix $h\in\cg$, we can find
coefficients $(a_\b)_{\b<0}$ such that $h-\sum_{\b<0}a_\b\bar
I_{\b,L}$ is in $\cg^{ex}$, where $\bar I_{\b,L}(\n)=|\L_L|^{-1}\sum_x
I_\b(\n_x)$. Therefore, it is enough to
prove the lemma in the case where $h\in\cg^{ex}$ and in the case where
$h=\bar I_{\b,L}(\n)$. The first case is a 
straightforward generalization of Lemma 4.4 in [CLO] since the
integration by parts formula is valid in $\cg^{ex}$.
In the second case, denote by $\hat I_\b(\n_x)$ ($\b<0$ fixed) the conditional
expectation of $I_\b(\n_x)$ with respect to the empirical averages of the
conserved quantities $\bar I^+_L$.  
%and let $\hat I^\ve_{\b,L}=|\L_L|^{-1}\sum_x\hat I_\b(\n_x)$.
Then the left hand side of \equ(apriori) is
bounded above (up to a factor 2) by the sum of the two terms
$$
\E^\mu_\ve\left[\sup_{0\le t\le T}\left(
\int_0^t\!ds\,\ve^{3/2-1}\bar G^B_s
\sum_x\bl(I_\b(\n_x)-\hat I_\b(\n_x)\br)\,ds\right)^2\right],
\Eq(bg5)
$$
where $\bar G^B_s:=\ve^3 \sum_{x\in B}G(s,\ve x)$, and 
$$
\E^\mu_\ve\left[\sup_{0\le t\le T}\left(
\int_0^t\!ds\,\ve^{3/2-1}\sum_{x\in B}G(s,\ve x)
\hat I_\b(\n_0)\,ds\right)^2\right],
\Eq(bg6)
$$
From the inequality
\equ(bg1), \equ(bg5) is less or equal to
$$
c\,V_L\bl(\tilde I_\b(\n_0),r,\t\br)\int_0^T\!ds\,\bl(\bar G^B_s\br)^2
$$
and by corollary 4.6 of [EMY2], 
$V\bl(\tilde I_\b(\n_0),r,\t\br)=\limsup_L
V_L\bl(\tilde I_\b(\n_0),r,\t\br)<+\infty$. So \equ(bg5) is bounded
above by
$$
c\int_0^T\!ds\,\ve^3\sum_{x\in B}G^2(s,\ve x)
$$
($c$ a positive constant).
Finally, by stationarity of $\mu$, the term \equ(bg6) is less than
$$
\ve^{-5}\E^\mu\bl[(\hat I_\b(\n_0))^2\br]T\int_0^T\!ds\,\bl(\bar
G^B_s\br)^2.
$$
From Corollary \equ(lbg0b), $\E^\mu\bl[(\hat
I_\b(\n_0))^2\br]\le c\,\ve^6$
and \equ(bg6) is going to zero as $\ve\to 0$.

\noindent\qed

\bigskip

Finally, Corollary \equ(lbg0b) and Lemma \equ(lbg1) allows to extend
straightforwardly the proof of the Boltzmann-Gibbs principle given in
section 4 of [CLO] and then to obtain Theorem \equ(s2.3).

\bigskip

We conclude this section by pointing out that the arguments for the
proof of tightness (section 5 of [CLO]) can be easily adapted 
to our case. 
Notice that, up to now, we did not need to have the
supremum over time inside the expectation in the Boltzmann-Gibbs
statement, however it is used in this part to control some terms
arising in martingale compensators. So, we can state

\medskip
\noindent{\bf \Theorem (tight)}
{\it
The family of probability $(Q^\ve)_{\ve>0}$ on $D([0,T],\ch_{-k_0})$
is tight since
$$
\lim_{M\to\infty}\lim_{\ve\to 0}\P^\mu_\ve\Bl(\sup_{0\le t\le T}
\|\xi_t^\ve\|_{-k_0}>M\Br)=0\Eq(t1)
$$
and for any $a>0$
$$
\lim_{\d\to 0}\lim_{\ve\to 0}\P^\mu_\ve\Bl(
\sup_{\scriptstyle |s-t|<\d\atop 0\le s,t\le T}
\|\xi^\ve_t-\xi^\ve_s\|_{-k_0}>a\Br)=0.\Eq(t2)
$$
}

\bigskip

\vskip 1truecm
\noindent{\bf  Appendix}
\vskip .5truecm \numfor=1
            
Let ${\cal A}$ be the space of $n\times n$ matrices with complex
entries.  ${\cal A}$ is a Hilbert space under the scalar product
$$
(X,Y)=\sum_{1\le k,\ell\le n}\bar X_{k,\ell}Y_{k,\ell}.
$$
Given a matrix $A$
in ${\cal A}$ the projector $\Pi_A $ is defined as the orthogonal 
projection onto ${\cal C}(A)$, the commutator space of $A$
$$
{\cal C}(A) =\{M\in A: [M,A]=0\},\quad [M,A]:=MA-AM.
$$
\vskip.2cm

\goodbreak
\noindent{\bf Lemma A.1}
\nobreak {\it Let $A$ be a diagonalizable matrix, ${\rm Sp}(A)\in i\R$.
  Then, for any matrix $M$ and $t>0$
  $$\lim_{\ve\to 0}{1\over t}\int_0^t\!ds\,\exp\bl({s\over\ve }A\br) M\exp
  \bl(-{s\over\ve }A\br)=\Pi_A(M). %\Eqa(A.1)
  $$
  }

\medskip

\noindent{\it Proof.} We follow the proof in [EP]. Let $P$ be a non-singular
matrix and $R$ a real diagonal matrix such that $A=P^{-1}i RP$. Let
$\{S_j,j=1,\dots,m\}$ be a partition of the integers
$\{j=1,\dots,n\}$ such that
$$
\eqalign{R_k&=R_\ell \quad \hbox{if} \quad k,\ell\in S_j \quad
  \hbox{for some}\ j,\cr R_k&\ne R_\ell\quad \hbox{otherwise,}}
$$
where $R_j, j=1,\cdots,n$ are the eigenvalues of $R$.  We define the
bar operation in the following way: let $K=(K_{k,\ell})\in {\cal A}$ be
$$K_{k,\ell}=\cases{1\quad \hbox{if} \quad k,\ell\in S_j \quad \hbox{for
    some}\ j,\cr 0 \quad \hbox{otherwise.}}
$$
Then $\ov M$, $M\in {\cal A}$, is defined as
$$\ov M_{k,\ell}=K_{k,\ell}M_{k,\ell}.$$
Observe that $\ov M$ is the diagonal
part of $ M$ in the simple case of $R$ with distinct eigenvalues.

\medskip

We have that
$$
\exp\bl({s\over\ve }A\br) M\exp \bl(-{s\over\ve }A\br)=
P^{-1}\exp\bl({i s\over\ve }R\br)PMP^{-1}\exp\bl(-{i s\over\ve }R\br)P.
$$
It is proved in [EP] that
$$
P^{-1}\overline{ P MP^{-1}} P
$$
is a projection onto ${\cal C}(A)$.
Hence $\ov M=\Pi_R(M)$ because $R$ is diagonal. Moreover,
$\Pi_R(M)=\Pi_A(M)$ because $R$ is diagonal. So it is enough to prove
that for any $M$
$$
\lim_{\ve\to 0}{1\over t}\int_0^t\!ds\,
\exp\bl({i s\over\ve }R\br) M\exp\bl(-{i
  s\over\ve }R\br)=\ov M.
$$
In [EP] it is also shown that for any matrix $M$
there exists a matrix $S$ such that $M$ can be decomposed as
$$
M=\ov M +[S,R].
$$
Since $\ov M$ commutes with $R$
$$
\exp\bl({i s\over\ve }R\br) M\exp\bl(-{i s\over\ve }R\br)
=\ov M +\exp\bl({i s\over\ve }R\br)[S,R]\exp\bl(-{i s\over\ve }R\br).
$$
The second term on the r.h.s gives
$$
\Bl(\exp\bl({i s\over\ve }R\br)[S,R]\exp\bl(-{i s\over\ve }R\br)\Br)_{k,\ell}=
S_{k,\ell}(R_\ell-R_k)\exp\bl({is\over\ve}(R_k-R_\ell)\br),
$$
where $S=(S_{k,\ell})$ and
$R=(R_{k,\ell})=(R_k\d_{k,\ell})$.  As a consequence,
$$
\lim_{\ve\to 0}{1\over t}\int_0^t\!ds\exp\bl({i s\over\ve
  }R\br)[S,R]\exp\bl(-{i s\over\ve }R\br)=0.
$$
\vskip.4cm 

\noindent{\bf Lemma A.2} {\it Let $\E$ be a first order differential
operator such that its Fourier transform $\hat\E(k)$ satisfies ${\rm
  Sp}(\hat\E(k))\in i\R$ for any $k$ and let
  $\D=\sum_{\a,\g=1}^3D_{\a,\g}\partial_\a\partial_\g$ be a second order
  differential operator, where $D=(D_{\a,\g})=(D^{\b,\nu}_{\a,\g})$ is
  a definite positive rank 2 tensor. 
  Then there exists a definite positive second order differential operator 
 $\pi_\E(\D)$ such that for any $G$ smooth
$$
\lim_{\ve\to 0}\Bigl\|\int_0^t\!ds\,\Big[\exp\bl({s\over\ve }\E\br) 
\D\exp \bl(-{s\over\ve }\E\br)-\pi_\E(\D)\Big]G\Bigr\|_0=0.
$$
}
\medskip

\noindent{\it Proof.} Let $\hat \D(k)$ be the Fourier transform of $\D$
$$
\hat\D(k)=-\sum_{\a,\g=1}^3 D_{\a,\g}^{\b,\nu}k_\a k_\g
\hat G(k)
$$
It is enough to prove that for any $t>0$ and for any $G$
smooth there exist a matrix $\hat\pi_E(\D)$ such that
$$
\lim_{\ve\to 0}\Bigl\|\int_0^t\!ds\,\Bigl[\exp\bl({s\over\ve }\hat \E\br)
\hat\D\exp\bl(-{s\over\ve }\hat \E\br)-\hat\pi_\E(\D)\Bigr] G\Bigr\|_0=0
$$
where $\|\cdot\|_0$ is the usual norm in $L^2(\T_3,\R^5)$. 
Choosing $\hat\pi_\E(\D)=\pi_{\hat\E}(\hat\D)$, 
that is an easy consequence of Lemma A.1 via 
dominated convergence theorem since, by assumption, $\hat\E$
is diagonalizable with pure complex eigenvalues which implies
$\bigl\|\exp \bigl({s\over\ve }\E\bigr)\bigr\|_0\le {\rm const}$.
Finally, since $\hat\D$ is definite positive, the same is true for 
$\pi_{\hat\E}(\hat\D)$.

\noindent\qed
\smallskip

Notice that Lemma A.2 implies that for any $0\le s\le t$,
$$
\lim_{\ve\to 0}\Bigl\|\int_s^t\!du\,\Big[\exp\bl({u\over\ve }\E\br) 
\D\exp \bl(-{u\over\ve }\E\br)-\pi_\E(\D)\Big]G\Bigr\|_0=0
$$

\vskip.4cm
\goodbreak

\noindent{\bf Lemma A.3}
\nobreak {\it Let $A^\ve(s)$, $A$ be linear operators from
$\ch_{k_0+2}$ to $\ch_{k_0}$ such that
$$
\sup_{\ve,0\le s\le t}\bl\|A^\ve(s)\br\|_{k_0+2\to k_0}<\infty
$$
\nobreak
and for any $G\in \ch_{k+2}$ and $0\le s< t$
  $$\lim_{\ve\to 0}\lt\|\int_s^t\!du\, [A^\e(u)-A]G\rt\|_{k_0}=0.$$
  Then, for any $G\in \ch_{k_0+2}$
  $$\lim_{\ve\to
    0}E_\ve^{\mu}\lt[\lt(\int_0^t\!ds\,\xi^\ve
  \bl(s,[A^\ve(s)-A]G\br)\rt)^2\rt]=0,
  $$
  where $\xi^\ve_s$ is the fluctuation field.}

\bigskip

\noindent{\it Proof.} We set $\ps{\xi^\ve_t,G}=\xi^\ve(t,G)$.
Let $0=t_0<t_1<\dots<t_\ell=t$ be a subdivision of the
interval $[0,t]$ of size $\d>0$. Then
$$
\int_0^t\!ds\,\xi^\ve(s,A^\ve(s)G)
=\sum_{i=0}^{\ell-1}\Bl\langle\xi^\ve_{t_i},\int_{t_i}^{t_i+1}\!ds\,A^\ve(s)G
\Br\rangle+R_1^\ve,
$$
with
$$R_1^\ve
  =\sum_{i=0}^{\ell-1}\int_{t_i}^{t_i+1}\!ds\,\ps{\xi_s^\ve
  -\xi_{t_i}^\e,A^\ve(s)G}.
$$
Since
$$
  |R_1^\ve|\le t\sup_{\scriptstyle |s_1-s_2|\le\delta\atop\scriptstyle
    0\le s_1,s_2\le t}
  \|\xi_{s_1}^\ve-\zeta_{s_2}^\e\|_{-k_0}
  \sup_{0\le
    s\le t}\|A^\e(s)\|_{k_0+2\to k_0}\|G\|_{k_0+2},
$$
it results from tightness \equ(t2) that for any $\d>0$,
$$
\lim_{\d\to 0}\lim_{\ve\to 0}P^\ve_{\mu}\bl(|R_1^\ve|>\d\br)=0.
$$
Moreover
$$\int_0^t\!ds\,\ps{\xi^\ve_s,A^\ve(s)G}=
\sum_{i=0}^{\ell-1}\ps{\xi_{t_i}^\ve,AG}(t_{i+1}-t_i)+R_1^\ve+R_2^\e,$$
with
$$
\eqalign{
|R_2^\e|&
=\left|\sum_{i=0}^{\ell-1}\int_{t_i}^{t_{i+1}}
\!ds\,\ps{\xi^\ve_{t_i}(A^\ve(s)-A)G} \right|\cr
&\le t\sup_{0\le s\le
    t}\|\zeta_s^\ve\|_{-k_0}\max_i\left\|\int_{t_i}^{t_{i+1}}ds[A^\ve(s)-A]G
\right\|_{k_0}.
}
$$
From assumption
$$
\lim_{\ve\to 0}\max_i\lt\|\int_{t_i}^{t_{i+1}}\!ds\,
[A^\ve(s)-A]G\rt\|_{k_0}=0.
$$
So, using tightness \equ(t1), we get for $M>0$ and $\ve$ small enough
$$
P^\ve_{\mu}\bl(|R_2^\ve|>\d\br)\le P^\ve_{\mu}\Bl(
\sup_{0\le s\le t}\|\xi_s^\e\|_{-k_0}>{\d\over M}\Br)
$$
which vanishes in the limit
${M\to0}$ after ${\ve\to 0}$.
With the same kind of arguments (using tightness again), we get
$$
\sum_{i=0}^{\ell-1}\ps{\xi_{t_i}^\ve,AG}(t_{i+1}-t_i)
=\int_0^t\!ds\,\ps{\xi_s^\e,AG}
+R_3^\ve,
$$
where 
$$
\lim_{\d\to 0}\lim_{\e\to 0}P^\ve_{\mu}\bl(|R_4^\e|>\d\br)=0.
$$
We have proved so far that $\int_0^t\!\,ds\bl(\xi^\ve(s,A^\ve(s)G)-
\xi^\ve(s,A G)\br)$
converges to 0 in $\P^\mu_\ve$ probability. To assert that the convergence
occurs in $L^2(P^\mu)$ it suffices e.g. to check that 
$$
\sup_\ve \E^\mu_\ve\lt[\lt(\int_0^t\!ds\,\ps{\xi^\ve_s,(A^\ve(s)-A)G}\rt)^4\rt]
<\infty,
$$
which is clear from the assumptions on the operator $A^\ve(s)$ and $A$.

\vfill\eject

{\centerline{\bf REFERENCES}} \vskip .4cm

\item {[BEM]} O. Benois, R. Esposito and R. Marra {\it Navier-Stokes
    limit for a thermal stochastic lattice gas.}  J. Stat Phys. {\bf
    90} 653--713 (1999).
  
\item{[BR]} T. Brox and H. Rost, {\it Equilibrium fluctuations of
    stochastic particle systems: the role of conserved quantities}.
  Ann. Probab. {\bf 12}, 742-759 (1984).

\item{[C1]} C.C. Chang, {\it Equilibrium fluctuations of gradient
    reversible particle systems}.  Probab. Th. Rel. Fields {\bf 100},
  269--283 (1994).
  
\item{[C2]} C.C. Chang, {\it Equilibrium fluctuations of nongradient
    reversible particle systems}.  In: Funaki, T., Woyczynski, W.A.
  (ed.): Nonlinear stochastic PDE's: Burgers turbulence and
  hydrodynamic limit, IMA volume {\bf 77}, pp. 41--51, Springer, (1996).
  
\item {[CLO]} C.C. Chang, C. Landim and S.Olla, {\it Equilibrium
    fluctuations of asymmetric simple exclusion processes}. To appear on
 Probab. Th. Rel. Field
  (2000).

\item{[CY]} C.C. Chang and H.T. Yau, {\it Fluctuations of one dimensional
Ginzburg-Landau models in Nonequilibrium} Commun. Math. Phys, {\bf 145}. 209-234
(1992).

\item{[EP]}S. Ellis and A. Pinsky {\it The projection of the
    Navier-Stokes Equations upon the Euler Equations} J.Math. Pures
  and Appl. {\bf 54}, 157--182 (1975).

%\item{[{EM}]}  
%R. Esposito and R. Marra,  
%{\it On the derivation of the incompressible Navier-Stokes equation
%for Hamiltonian particle systems}, 
%J. Stat. Phys., {\bf  74}, 981-1004 (1993).
%
%\item{[{EMY1}]}
%R. Esposito, R. Marra and H.T. Yau, 
%{\it Diffusive limit of asymmetric  simple exclusion},
%Review in Math. Phys. {\bf 6}, 1233-1267 (1994).
%
%\item{[{EMY2}]}
%R. Esposito, R. Marra and H.T. Yau,
%{\it Diffusive limit of the asymmetric simple exclusion: the Navier-Stokes
%correction},
%Nato ASI Series B: Physics {\bf 324}, 43--53 
%{\it On three levels:
%Micro-Meso and Macro- Approaches in Physics},
%M. Fannes-C. Maes-A. Verbeure edt.s
%(1994).
%
%\item{[EM]} R. Esposito and R. Marra {\it Hydrodynamics as scaling
%     limit of kinetic systems and stochastic particle systems.}  to
%   appear on Research Trends in Physics, (2000).
% 
\item{[EMY1]}
R. Esposito, R. Marra and H.T. Yau, 
{\it Diffusive limit of asymmetric  simple exclusion},
Review in Math. Phys. {\bf 6}, 1233-1267 (1994).

\item{[{EMY2}]} R.Esposito, R.Marra and H.T. Yau, {\it Navier-Stokes
    equations for stochastic particle systems on the lattice}, Commun.
  Math.  Phys. {\bf 182}, 395--456 (1996).
  
\item{[FF]} Ferrari, P.A., Fontes, L.R.G.: Shock fluctuations in the
  asymmetric simple exclusion process. Probab. Th. Rel. Fields {\bf
    99}, 305--319 (1994).
  
\item{[GP]} G{\"a}rtner, J., Presutti, E.: Shock fluctuations in a
  particle system. Ann. Inst. H. Poincar\'e, Physique Th\'eorique {\bf
    53}, 1--14 (1990).
%  
%\item{[Fr]} A. Friedman {\it Partial Differential Equations of
%    Parabolic Type}, Prentice-Hall, Englewood Cliffs, N. J. (1964 ).
% 
%\item{[GPV]} M. Guo, G.C. Papanicolau and S.R.S. Varadhan, {\it Non
%    linear diffusion limit for a system with nearest neighbor
%    interactions}, Comm. Math. Phys. {\bf 118}, 31--59 (1988).
  
\item{[HS]} R.A. Holley and D.W. Strook {\it Generalized
    Ornstein-Uhlenbeck processes and infinite branching Brownian
    motions}. Kyoto Univ. RIMS {\bf 14}, 741--814 (1978).
  
\item{[KL]} C.  Kipnis and C. Landim, {\sl Hydrodynamic limit of
    interacting particle systems}, Springer-Verlag, (1999).
  
\item{[L]} S.L. Lu, {\it Equilibrium fluctuations of a one dimensional
    nongradient Ginzburg-Landau model}. Ann. Probab. {\bf 22},
  1252-1272 (1994).
%  
%\item{[LY]} C. Landim and H.T. Yau, {\it Fluctuation--dissipation
%    equation of asymmetric simple exclusion processes}, Probab. Theory
%  Relat. Fields {\bf 108}, 321--356 (1997).
%
%  
%\item{[QY]} J. Quastel and H.T. Yau {\it Lattice gases and the
%    incompressible Navier-stokes equations}, preprint, (1988).
%  
\item{[S]} H. Spohn, {\it Large Scale Dynamics of Interacting
    Particles}, Springer-Verlag, New York (1991).
%
%  
%\item{[V]} S. R. S. Varadhan, {\it Nonlinear diffusion limit for a
%    system with nearest neighbor interactions II}, in {\it Asymptotic
%    Problems in Probability Theory: Stochastic Models and Diffusion on
%    Fractals}, K.D. Elworthy and N. Ikeda eds, Pitman Research Notes
%  in Math., {\bf 283}, 75--128 J. Wiley \& Sons, New York (1994).

\bigskip

\noindent{\bf Acknowledgments}: Two of us (R.E. and R.M.) wish to 
thank the University
of Rouen and the IHES in Bures-sur-Yvette, where part of this work was
done, for the very kind hospitality. Work partially supported by
GNFM-INDAM, MURST and CEE-TMR's on Hyperbolic Systems and Kinetic
Models. 

O.B. acknowledges very much the University of Roma Tor Vergata for hospitality,
C. Landim for fruitful discussions and is infinitely grateful to his
two co-authors for them patience and understanding about the
difficulty to become a father.

\end